\documentclass[aps,pra,showpacs,amsmath,amssymb,reprint,10pt,longbibliography,noeprint]{revtex4-1}
\pdfoutput=1

\usepackage{amsmath,amsfonts,amssymb,amsthm,bbm,graphicx,enumerate,times}
\usepackage{mathtools}
\usepackage[usenames,dvipsnames]{color}
\usepackage{todonotes} %[disable]
\usepackage{comment}
\usepackage{float}
\usepackage{listings}
\usepackage{hyperref}
\usepackage{tikz}
\usepackage{graphicx}
\usepackage{bm}
\usetikzlibrary{shapes,positioning}
\usepackage{etoolbox}

 \definecolor{jens}{rgb}{.1,0.5,.4}

 \newtheorem{theorem}{Theorem}

 \newtheorem{lemma}[theorem]{Lemma}
 \newtheorem{corollary}[theorem]{Corollary}
 \newtheorem{definition}[theorem]{Definition}

\definecolor{henrik}{rgb}{.8,.3,0}

\newcommand{\mc}[1]{\mathcal{#1}}

\newcommand{\mb}[1]{\mathbb{#1}}

\newcommand{\e}{\mathrm{e}}

\newcommand{\tr}{\mathrm{Tr}} %old
\newcommand{\Tr}{\mathrm{Tr}} %new

\newcommand{\id}{\mb{1}}

\newcommand{\1}{\mathrm{id}}

\renewcommand{\1}{\id}

\newcommand{\ket}[1]{\left.\left|{#1}\right.\right\rangle}

\newcommand{\bra}[1]{\left.\left\langle{#1}\right.\right|}

\newcommand{\ketbra}[2]{\ket{#1} \!\! \bra{#2}}

  \newcommand{\proj}[1]{\ketbra{#1}{#1}}
\renewcommand{\vec}[1]{\pmb{#1}}

\begin{document}
\title{Entropy and reversible catalysis}
 
\author{H.\ Wilming}
%%\email{henrikw@phys.ethz.ch}
\affiliation{Institute  for  Theoretical  Physics,  ETH  Zurich,  8093  Zurich,  Switzerland,\\
Leibniz Universit\"at Hannover, Appelstra\ss e 2, 30167 Hannover, Germany}
%\author{P.\ Boes}
%\affiliation{\fu}
\begin{abstract}
I show that non-decreasing entropy provides a necessary and sufficient condition to convert the state of a physical system into a different state by a reversible transformation that acts on the system of interest and a further "catalyst" whose state has to remain invariant exactly in the transition.
This statement is proven both in the case of finite-dimensional quantum mechanics, where von~Neumann entropy is the relevant entropy, and in the case 
of systems whose states are described by probability distributions on finite sample spaces, where Shannon entropy is the relevant entropy.
The results  give an affirmative resolution to the (approximate) "catalytic entropy conjecture" introduced by Boes et al. [PRL 122, 210402 (2019)].
They provide a complete single-shot characterization without external randomness of von Neumann entropy and Shannon entropy. 
	I also compare the results to the setting of phenomenological thermodynamics and show how they can be used to obtain a quantitative single-shot characterization of Gibbs states in quantum statistical mechanics.
\end{abstract}
\maketitle
A central question in quantum information theory is which quantum states on some physical system may be transformed into which other states on the same (or a different) physical system by a given set
of operations. 
This question underlies quantum resource theories, such as entanglement \cite{Bennett1996,Bennett1996a,Vedral1997,Horodecki2009}, thermodynamics \cite{Lieb1999,Janzing2000,Horodecki2013,Brandao2013,Brandao2015,Lostaglio2019} or asymmetry \cite{Bartlett2007,Marvian2014,Marvian2014a} (see Ref.~\cite{Chitambar2019} for a review on quantum resource theories).
Common to most resource theories is that they allow for probabilistic mixing of operations, i.e., to use a source of classical randomness (such as a coin toss) to decide on the operation that is implemented.
This use of randomness immediately implies that the resource theory is \emph{convex}, which greatly simplifies the mathematical analysis.
While often a natural assumption due to the commonplace access to (quasi-)randomness and classical communication, one may ask what happens if either one considers the cost of classical or quantum randomness  explicitly or simply disallows the use of classical or quantum randomness.
In the most extreme limit one would then end up only allowing the use of unitary operations and the question of which states can be inter-converted becomes trivial: namely all those states which are unitarily invariant, or in other words, all states with the same spectrum (including multiplicities).

In many resource theories the use of a \emph{catalyst} is operationally well motivated and  may greatly enrich the set of possible state transitions 
\cite{Jonathan1999,Daftuar2001,Dam2003,Aubrun2007,Klimesh2007,Turgut2007,Brandao2015,Ng2015,Boes2018b}. A catalyst is a system which remains invariant in a given process, but may or may not, depending on the resource theory, build up correlations to other systems. Since its state does not change in a process, no resources are used up and the catalyst may be used again to facilitate further state transitions on other systems -- this is conceptually similar to catalysts in chemistry or the ubiquitous "periodically working machines" in thermodynamics. 
One may therefore wonder what happens if we consider only unitary operations together with the possibility for catalysts.  
Interestingly, Ref.~\cite{Boes2018b} found that in this setting the von~Neumann entropy plays a special role due to its sub-additivity property (see below). 
It was in fact conjectured that the von Neumann entropy is the \emph{only} constraint for state-transitions once one allows for arbitrary small errors on the system (but not the catalyst). 
This conjecture was called "catalytic entropy conjecture" and can also be formulated in the classical setting as a conjecture connecting Shannon entropy to catalytic permutations of probability distributions. 
The ideas behind this conjecture already have found application in the context of thermodynamics and fluctuation theorems \cite{Sparaciari2017,Boes2020a}. 

The special role of von Neumann entropy is surprising for several reasons: Typically, the Shannon or von Neumann entropy appears in settings involving many weakly correlated systems due to the phenomenon of \emph{typicality} \cite{Shannon1948,Cover2006,Wilde2009}. 
It was therefore long believed that the von Neumann entropy only plays a special role in asymptotic settings, such as the thermodynamic limit in physics or the limit of many identically and independently distributed signals in information theory. 
In particular, in resource theories involving free randomness and allowing for catalysts (that however may not become correlated to the system of interest),  state transitions are usually characterized by an infinite set of constraints \cite{Aubrun2007,Klimesh2007,Turgut2007,Brandao2015}. 
It was only very recently first conjectured and then proven that these conditions may collapse to the von Neumann entropy (or similar quantities, such as relative entropy or free energy) if one allows for catalysts that become correlated to the system \cite{Gallego2015,Wilming2017a,Lostaglio2015b,Mueller2016,Mueller2017,Rethinasamy2020,Shiraishi2020}. 
However, these settings still made use of external randomness -- either by allowing for classical randomness  explicitly or allowing free access to systems such as heat baths, which can be seen as sources of randomness. 
It is thus interesting that the catalytic entropy conjecture posits that von Neumann entropy plays such a special role in situations that neither allow for the use of external randomness, nor require an asymptotic limit. 
This paper provides an affirmative resolution of the catalytic entropy conjecture.

%%%%%%%%%%%%%%%%%%%%%%
\begin{figure}[t!]
		\includegraphics[width=7cm]{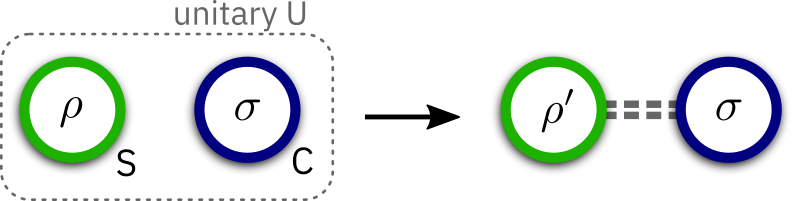}
	\caption{A catalytic transition: A unitary operation $U$ is applied to systems $S$ and $C$ in the state $\rho\otimes\sigma$. The resulting reduced state on $S$ is (arbitrarily close to) $\rho'$, while the reduced state on $C$ is preserved exactly, but correlated to $S$ (indicated by the dashed lines).
	The main result of this paper shows that a matching $\sigma$ and $U$ can be found if and only if $H(\rho')\geq H(\rho)$. }
		\label{fig:catalytic_transition}
\end{figure}
%%%%%%%%%%%%%%%%%%%%%5

The catalytic entropy conjecture can also be formulated in full analogy in the classical case by replacing finite dimensional density matrices with probability distributions on finite sample spaces, unitary operations with permutations (reversible transformations on the sample space) and von Neumann entropy with Shannon entropy. 
In the main text of this paper we will only consider the quantum case and only briefly comment on how to proof the classical version of our main result. 
A full proof of the classical result is given in Appendix~\ref{appendix:classical}. While the proof of the classical result also implies the quantum result, it has the drawback of requiring a larger catalyst in general.

\emph{Setting and main result.} Throughout we consider a system $S$ described by density matrices $\rho$ and $\rho'$ on a Hilbert-space of dimension $d$. 
In the following we write $H(\rho)$ for von Neumann entropy, which is defined as
\begin{align}
	H(\rho) = -\tr[\rho \log(\rho)]. 
\end{align}
Von Neumann entropy is continuous in $\rho$ \cite{Fannes1973,Audenaert2007} and has several useful properties, such as unitary invariance ($H(\rho) = H(U\rho U^\dagger)$ for any unitary $U$), additivity ($H(\rho\otimes \sigma) = H(\rho)+H(\sigma)$)
and sub-additvity: $H(\rho_{12})\leq H(\rho_1) + H(\rho_2)$, where $\rho_{12}$ is a bipartite quantum state with marginals $\rho_1$ and $\rho_2$. Finally, denote by
$D(\rho,\rho') := \frac{1}{2} \| \rho-\rho' \|_1$ the trace-distance between two density-matrices. 
We then define catalytic state transitions formally as follows (see also Fig.~\ref{fig:catalytic_transition}).
\begin{definition}[Approximate catalytic transformation]
Consider two finite-dimensional density matrices $\rho$ and $\rho'$ on the same system $S$. 
We write $\rho\rightarrow_\epsilon \rho'$ if there exists a finite-dimensional density matrix $\sigma$ on a system $C$ and a unitary $U$ on $SC$ such that 
	\begin{align}
		\Tr_S[U\rho\otimes \sigma U^\dagger]=\sigma
	\end{align}
and 
	\begin{align}
		D\big(\Tr_C[U\rho\otimes\sigma U^\dagger],\rho'\big) \leq \epsilon.
	\end{align}
\end{definition}
The following main result of this paper then shows that the set of states that are reachable from a given state $\rho$ is given exactly by the set of states with higher von Neumann entropy:
\begin{theorem}[Catalytic transformations characterize von Neumann entropy]\label{thm:cec}
The following are equivalent:
	\begin{enumerate}[i)]
		\item $\rho\rightarrow_\epsilon \rho'$ for all $\epsilon>0$.
		\item $H(\rho') \geq H(\rho)$.
	\end{enumerate}
\end{theorem}
%An isolated (quantum) system that undergoes (micro-)reversible dynamics has constant entropy. The result shows that if a system plus its environment undergo reversible dynamics
%in such a way that the (statistical) state of the environment remains unchanged at the end of the process, then the systems entropy is non-decreasing. 
%Moreover, \emph{any state} with higher entropy may be reached by a suitable environment and reversible dynamics.  
Before coming to the proof of this statement, let us first discuss the formal similarity between the theorem and the corresponding set-up in thermodynamics and then give an application for quantum statistical mechanics.

{\emph{Analogy with Thermodynamics. } Let us consider the following idealized, but ubiquitous setting of a thermodynamic work-process acting on three systems: A system $S$ composed of a working substance and all other parts
that may change in the process (such as heat baths), a collection of systems $C$ composed of all systems that take part in the process but return to their initial state when the process has finished (gears, pistons etc.), and finally an idealized work-storage device $W$ with vanishing entropy (e.g., a purely mechanical device such as a suspended weight in a uniform gravitational field).
Since the system $C$ is cyclic, it does not contribute any energy or entropy to the process. 
Similarly, $W$ only contributes or absorbs the work necessary for the process, but does not act as a source or sink of entropy. 
Let the initial and final states of $S$ be labelled by $a$ and $b$, respectively. 
By the Second Law of Thermodynamics, the considered process can only be possible if the entropy of $S$ is non-decreasing: $H(b) \geq H(a)$ (we use the same letter for thermodynamic entropy as for von~Neumann or Shannon entropy). 
Conversely, and importantly, however, it is commonly assumed in setting up the thermodynamic framework that for any two states $a,b$ of $S$ there is some such idealized work-process connecting the two states  (for a very clear, recent exposition see \cite{Kammerlander2020}). 
Indeed, this is required to be able to consistently define a thermodynamic entropy. 
Therefore, the states $a$ and $b$ can be connected by such a process if and only if $H(b)\geq H(a)$, in complete analogy with Theorem~\ref{thm:cec}, where the role of $W$ is played by the external laboratory implementing the unitary $U$. In the thermodynamic setting, the system $C$ is usually left implicit, but is clearly physically necessary and its precise design depends on the states $a$ and $b$. 
While in the thermodynamic case the source of the increase of entropy is left unspecified, in the microscopic case discussed here it is due to the build up of correlations between $S$ and $C$.
Explicitly, we have
%\begin{align}
$H(\rho') - H(\rho) = I(S':C)$,
%\end{align}
where $I(S':C)$ denotes the mutual information between $S$ and $C$ after the unitary evolution. 
} 

{\emph{Quantitative single-shot characterization of Gibbs states.} 
Let us now illustrate the use of the main result with an application in quantum statistical mechanics. We first have to establish some background material.
A \emph{passive state} is any quantum state $\rho$ whose energy (w.r.t. some fixed Hamiltonian $\mathsf H$) cannot be lowered by a unitary transformation. 
Equivalently, we may say that no work can be extracted from a passive state using unitary operations. 
Even though $\omega$ is passive, it is not necessarily \emph{completely passive}, meaning that a positive amount of energy
\begin{align}
	\overline W(\rho) := \lim_{n\to\infty} \frac{1}{n} \sup_{U}\left[\Tr[\rho^{\otimes n} \hat{\mathsf H}_n] - \Tr[U \rho^{\otimes n} U^\dagger \hat {\mathsf H}_n]\right]\nonumber 
\end{align}
may be extracted \emph{per copy} if many copies of $\rho$ are available. Here, $\hat{\mathsf H}_n := \sum_{j=1}^n \mathsf H_j$ with $\mathsf H_j\equiv \mathsf H$ is the total Hamiltonian on $n$ copies. 
Let us assume for simplicity that the ground state of $\mathsf H$ is unique. 
It has been proven that \cite{Alicki2013}
\begin{align}
	\overline W(\rho) = \Tr[\rho\,\mathsf H] - \Tr[\omega_{\beta(\rho)}(\mathsf H)\,\mathsf H],
\end{align}
where $\omega_\beta(\mathsf H) := \exp(-\beta \mathsf H)/\Tr[\exp(-\beta \mathsf H)]$ denotes a Gibbs state and $\beta(\rho) \geq 0 $ is chosen such that
$H(\omega_{\beta(\rho)}(\mathsf H)) = H(\rho)$ (if $H(\rho)=0$ one has to take the limit $\beta\rightarrow +\infty$). Thus, the only completely passive states are ground states and Gibbs states with positive temperature \cite{Pusz1978,Lenard1978}.

The above characterization of Gibbs states relies on a thermodynamic limit.
It was observed in Ref.~\cite{Sparaciari2017}, and discussed in detail for the particular case of three-level systems, that the energy of passive states may nevertheless in general be reduced using catalytic transitions instead of unitary operations unless the state in question is a Gibbs state. This may be surprising, since the catalyst, by virtue of not changing its state, cannot compensate for energetic changes. It was an open problem to determine how much work can be extracted from an arbitrary state using a single catalytic transformation \cite{footnotepassive}. Let us denote this quantity by
\begin{align}
	W_{\mathrm{cat.}}(\rho) := \sup_{\rho\rightarrow_{0} \rho'}\left(\Tr[\rho \mathsf H] - \Tr[\rho' \mathsf H]\right).
\end{align}
Then Theorem~\ref{thm:cec} immediately leads to the following corollary, which shows that catalysts allow to extract the same amount of energy from a single passive states as on average from asymptotically many copies:
\begin{corollary}\label{cor:passive}
		$W_{\mathrm{cat.}}(\rho) = \overline W(\rho).$
\begin{proof}
	By Theorem~\ref{thm:cec} and continuity we have 
	\begin{align}
	W_{\mathrm{cat.}}(\rho) = \sup_{\rho': H(\rho')\geq H(\rho)} \left(\Tr[\rho \mathsf H]-\Tr[\rho' \mathsf H]\right).
	\end{align}
	By concavity of von~Neumann entropy, the optimizer must have $H(\rho')=H(\rho)$. But by Gibbs variational principle, $\omega_{\beta(\rho)}(\mathsf H)$ is the state with lowest energy among all states with entropy $H(\rho)$.
	Hence $W_{\mathrm{cat.}}(\rho) = \Tr[\rho \mathsf H] - \Tr[\omega_{\beta(\rho)}(\mathsf H) \mathsf H] = \overline W(\rho)$. 
\end{proof}
\end{corollary}
}
\emph{Proof of the Theorem~\ref{thm:cec}.} 
The direction i) $\Rightarrow$ ii) follows directly from sub-additivity, unitary invariance and continuity of von~Neumann entropy: For a given $\epsilon$, denote by $\rho'_\epsilon$ the final state on $S$. Then we have
	\begin{align}
		H(\rho'_\epsilon) + H(\sigma) \geq H(U(\rho\otimes \sigma)U^\dagger) = H(\rho) + H(\sigma)
	\end{align}
	and hence $H(\rho'_\epsilon)\geq H(\rho)$. By continuity we thus find $H(\rho')\geq H(\rho)$. 

The converse direction ii) $\Rightarrow$ i) requires several Lemmas.
First, we collect a combination of some standard results on typicality and majorization.
We write $\rho\succeq \rho'$ if $\rho$ \emph{majorizes} $\rho'$, meaning that there exists a probability distribution $q_i$ over unitaries $V_i$ such that $\rho' = \sum_i q_i V_i \rho V_i^\dagger$.
Similarly, we write $\vec a \succeq \vec b$ for two vectors $\vec a,\vec b\in \mathbb R^d$ if there exists a probability distribution $q_i$ over permutation-matrices $\pi_i$ such that $\vec a = \sum_i q_i \pi_i\vec b$. 
\begin{lemma}[Typicality and majorization]\label{lemma:typmaj}
	Let $\rho$ and $\rho'$ be two finite-dimensional density matrices of dimension $d$ with $H(\rho) < H(\rho')$. 
	Then for any $\epsilon>0$ and large enough $n$ there exists a state $\rho'_{\epsilon,n}$ such that $\rho^{\otimes n} \succeq \rho'_{\epsilon,n}$ and $D(\rho'_{\epsilon,n},{\rho'}^{\otimes n})\leq \epsilon$. Moreover, the error $\epsilon$ can be bounded as 
	\begin{align}
		\epsilon \leq O(\exp(-n \Delta H^2/4))
	\end{align}
	with $\Delta H := H(\rho')-H(\rho)$ {and $\rho'_{\epsilon,n}$ may be chosen to have the same eigenbasis as $\rho'^{\otimes n}$}.
\end{lemma}
A proof-sketch of Lemma~\ref{lemma:typmaj} is given in Appendix~\ref{appendix:lemma4}. 
It is clear that the given error bound is not optimal for every choice of $\rho$ and $\rho'$, since, for example, $\epsilon=0$ is possible if $\rho \succeq \rho'$. However, Ref.~\cite{Holenstein2011} shows that the given error bound is essentially optimal up to constants as a bound that does not take into account detailed information about $\rho$ and $\rho'$. In Appendix~\ref{sec:size} we use this to give an estimate of the size of the catalyst.

The next Lemma will be essential to construct a candidate catalyst by making use of Lemma~\ref{lemma:typmaj}.  
It is based on the \emph{Schur-Horn theorem}, which states that for any $d\times d$ Hermitian matrix its vector of eigenvalues $\vec \lambda$ majorizes the vector of diagonal elements in every orthonormal basis. Conversely, every vector that is majorized by $\vec \lambda$ may be obtained as the diagonal elements in a suitable orthonormal basis. {In particular  if $\vec p$ and $\vec p'$ denote the ordered vectors of eigenvalues of two density matrices $\rho$ and $\rho'$, respectively, then $\rho\succeq \rho'$ if and only if $\vec p\succeq \vec p'$.}
In the following we denote by $\mc D_{\rho'}$  the \emph{dephasing channel} in the eigenbasis of $\rho'$ that acts as
\begin{align}
	\mc D_{\rho'}[\rho] = \sum_i \proj{i}\!\rho\!\proj{i},
\end{align}
where the $\ket{i}$ constitute an orthonormal eigenbasis of $\rho'$.

\begin{figure}[t!]
	\includegraphics[width=8.5cm]{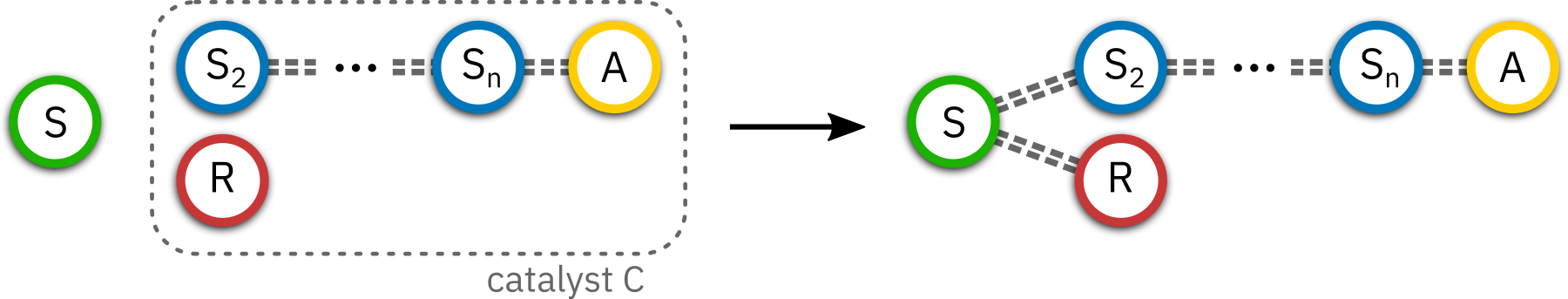}
	\caption{The structure of the constructed catalyst $C$: It contains subsystems $S_2,\ldots, S_n$, which are copies of the target system $S$ together with an auxiliary system $A$ of dimension $n$ as well as a catalytic source of randomness $R$. The dashed lines indicate possible correlations. The source of randomness is utilized to dilate the decoherence-channel $\mc D_{\rho'}$ on $S$ in such a way that it does not become correlated to the systems $S_2,\ldots,S_n$ and $A$ in the process.}
		\label{fig:structure_catalyst_q}
\end{figure}

\begin{figure*}[t!]
	\includegraphics[width=12cm]{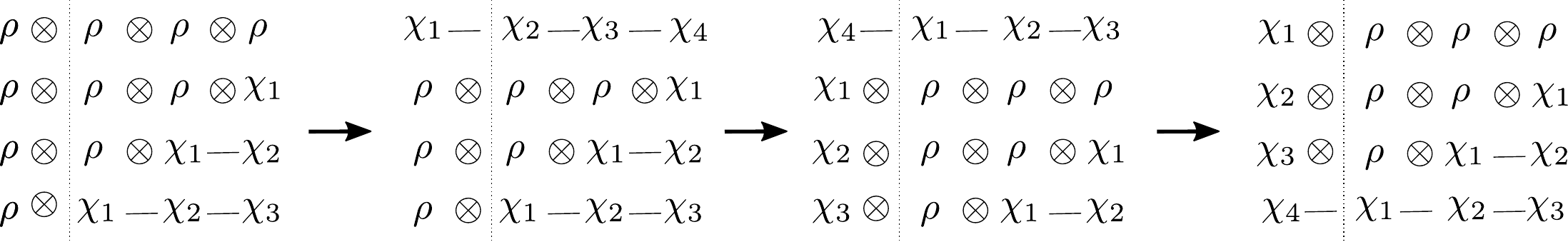}
	\caption{Illustration of the steps in the first part of the proof for $n=4$. The different columns denote the systems $S_1\cdots S_n$ with $S_1=S$ and the different rows indicate the state $\ket{k}$ of the auxiliary system $A$. The hyphen indicates that the corresponding subsystems may be correlated. This procedure is essentially identical to the one in Ref.~\cite{Shiraishi2020}. In the proof of the classical result (see Appendix~\ref{appendix:classical}), each of the subsystems with marginals called $\chi_i$ here is further correlated to the second part $R$ of the catalyst which is not shown.}
		\label{fig:proof}
\end{figure*}

\begin{lemma}[Basic lemma]\label{lemma:basic} Let $H(\rho)<H(\rho')$ and $\mc D_{\rho'}$ the dephasing channel in the eigenbasis of $\rho'$. Then for any $\epsilon>0$ there exists an $n\in\mathbb{N}$ and a unitary $U$ such that for any $1\leq k\leq n$:
	\begin{align}
		D\big(\rho',\mc D_{\rho'}[\chi_k]\big) \leq \epsilon,\quad \chi:= U \rho^{\otimes n} U^\dagger,
	\end{align}
	where $\chi_k := \Tr_{\{1,\ldots,n\}\setminus \{k\}}[\chi]$. The error $\epsilon$ scales as in Lemma~\ref{lemma:typmaj}.
	\begin{proof}
		We make use of the state $\rho'_{\epsilon,n}$ guaranteed by Lemma~\ref{lemma:typmaj}. 
		By the Schur-Horn theorem  there exists a unitary $U$ such that
		\begin{align}
			\rho'_{\epsilon,n}=\mc D_{\rho'}^{\otimes n}[\chi].
		\end{align}
		However, we have that (writing $\overline k := \{1,\ldots,n\}\setminus\{k\}$) 
		\begin{align}
			\mc D_{\rho'}[\chi_k]&=\tr_{\overline{k}}[(\mc D_{\rho'}\otimes \mathbf 1_{\overline{k}}[\chi]] = \tr_{\overline k}[\mc D_{\rho'}^{\otimes n}[\chi]] = \tr_{\overline k}[\rho'_{\epsilon,n}] \nonumber
		\end{align}
		by locality of quantum mechanics. But since the trace-distance is non-increasing under partial traces, we then find
		\begin{align}
			D\big(\rho',\mc D_{\rho'}[\chi_k]\big) \leq D\big({\rho'}^{\otimes n},\rho'_{\epsilon,n}\big)\leq \epsilon.
		\end{align}
	\end{proof}
\end{lemma}
The construction of the unitary $U$ guaranteed by the Schur-Horn theorem is explained in Appendix~\ref{appendix:howto}.
The final Lemma that we require provides a way for us to get rid of unwanted coherences (arising from Lemma~\ref{lemma:basic}) in the final state without correlating the catalyst internally (which would spoil the catalyst). 
\begin{lemma}[No propagation of correlations for mixed unitary channels]\label{lemma:mixed_unitary} 
	Consider a mixed unitary quantum channel $\mc C[\cdot] = \sum_i p_i V_i\,\cdot V_i^\dagger$ acting on a system $S$, where the $p_i$ denote probabilities and the $V_i$ are unitary operators.
	 Dilate $\mc C$ using an auxiliary system $C$ and state $\sigma = \sum_i p_i \proj{i}$ using the unitary $V:=\sum_i V_i\otimes\proj{i}$ as
	\begin{align}
		\mc C[\rho] = \Tr_2\left[V\rho\otimes \sigma V^\dagger\right].
	\end{align}
	Finally apply the dilation to a state $\rho_{S \overline{S}}$ on $S$ and a further system $\overline{S}$. Then
	\begin{align}
		\Tr_S\left[(V\otimes \mathbf 1_{\overline S})\rho_{S\overline S}\otimes \sigma (V^\dagger\otimes \mathbf 1_{\overline S})\right] = \rho_{\overline S}\otimes \sigma.
	\end{align}
	That is, the dilating system $C$ is catalytic and remains uncorrelated to $\overline{S}$.
	\begin{proof}
		The result immediately from the unitary invariance of the (partial) trace:
	\begin{align}
		\Tr_S&\left[(V\otimes \mathbf 1_{\overline S})\rho_{S\overline S}\otimes \sigma (V^\dagger\otimes \mathbf 1_{\overline S})\right]\\
		&= \sum_i p_i \Tr_S\left[(V_i\otimes \mathbf 1_{\overline S})\rho_{S\overline S}(V_i^\dagger\otimes \mathbf 1_{\overline S})\right] \otimes\proj{i} \nonumber\\
		&= \sum_i p_i \rho_{\overline S}\otimes\proj{i} = \rho_{\overline S}\otimes\sigma. 
	\end{align}
	\end{proof}
\end{lemma}
We are now in position to prove ii) $\Rightarrow$ i) of Theorem~\ref{thm:cec}. The proof proceeds in two parts. First we construct a catalyst $\sigma_1$ for the exact transition from $\rho$ to the equal mixture 
\begin{align}
\overline{\chi} := \frac{1}{n} \sum_{k=1}^n\chi_k 
\end{align}
of the states $\chi_k = \tr_{\overline k}[\chi]$, where $\chi$ is the state from Lemma~\ref{lemma:basic}. Then we use a second catalyst $R$ in state $\sigma_2$ to implement the dephasing map and obtain $\mc D_{\rho'}[\overline\chi]$ 
, which is $\epsilon$-close to the target $\rho'$. The part $R$ of the catalyst thus effectively acts as a source of randomness. By Lemma~\ref{lemma:mixed_unitary} and the fact that the dephasing map is a mixed unitary channel, this second part can be done in such a way that the two parts of the catalyst remain uncorrelated. 
{Therefore, the two-step process is still catalytic when both parts of the catalyst are considered as one joint catalyst in state $\sigma_1\otimes\sigma_2$. As a side comment, we mention that
the results of \cite{Boes2018a} imply that $\sigma_2$ only needs to have a dimension of the order of $\sqrt{d}$.}  
Furthermore, note that by perturbing $\rho'$ arbitrarily slightly, we can always ensure that $H(\rho)<H(\rho')$ since we allow for arbitrarily small errors and von Neumann entropy is continuous. 
We thus only need to prove that we can do the transition $\rho\rightarrow_{\epsilon=0} \overline{\chi}$ in the case $H(\rho')>H(\rho)$. To show this, we make use of a trick that was used in recent work by Shiraishi and Sagawa \cite{Shiraishi2020}:
We denote by $S=S_1$ the system and by $S_2,\ldots,S_n$ and $A$ subsystems of the first part of the catalyst. The $S_i$ all have the same Hilbert-space dimension as $S$ and $A$ has Hilbert-space dimension $n$ (see Fig~\ref{fig:structure_catalyst_q} for an illustration of the structure of the catalyst). 
Then define 
\begin{align}
	\sigma_1 = \frac{1}{n} \sum_{k=1}^n \rho^{\otimes k-1 }\otimes \chi_{1,\ldots,n-k}\otimes \proj{k}_A,
\end{align}
where $\chi_{1,\ldots,i}$ denotes the reduced density matrix of $\chi$ consisting of the subsystems $1$ to $i$ and we define $\chi_0$ and $\rho^{\otimes 0}$ to be the trivial state $1$.
We now apply the following sequence of unitaries on $\rho\otimes \sigma_1$ (see Fig.~\ref{fig:proof}):
\begin{enumerate}
	\item  $U\otimes\proj{n} + \mathbf 1\otimes\sum_{k=1}^{n-1}\proj{k}$, with $U$ the unitary from Lemma~\ref{lemma:basic}.
\item The cyclic shift of sub-systems $S_i \rightarrow S_{i+1}$ with $S_n\rightarrow S_1$. 
\item The cyclic shift on $A$, acting as $\ket{i}\rightarrow \ket{i+1}$ with $\ket{n+1}=\ket{1}$.
\end{enumerate}
After the three steps the catalyst is back to its initial state. The state on the system, on the other hand, is given by $\overline{\chi}$ and, after applying the dephasing map using the system $R$, we find
\begin{align}
	D\big(\frac{1}{n}\sum_{k=1}^n\mc D_{\rho'}[\chi_k],\rho'\big) \leq \frac{1}{n}\sum_{k=1}^n D\!\left(\mc D_{\rho'}[\chi_k],\rho'\right) \leq \epsilon\nonumber
\end{align}
by {the triangle inequality} and Lemma~\ref{lemma:basic}. %This finishes the proof.

%%%%%%%%%%%%%%%%%%%%%%
%%%%%%%%%%%%%%%%%%%%%5

%%%%%%%%%%%%%%%%%%%%%%
%%%%%%%%%%%%%%%%%%%%%5

\emph{The classical case.} As mentioned above, we can also formulate a the classical version of Theorem~\ref{thm:cec}.
To do that, we can define a catalytic transition $\vec p \rightarrow_\epsilon \vec p'$ between two probability vectors $\vec p,\vec p'\in \mathbb R^d$ as in the quantum case,
but with the catalysts' density matrix replaced by a further probability vector $\vec q\in \mathbb R^{d_C}$ and the unitary $U$ replaced by a permutation acting on the canonical basis vectors of $\mathbb R^d\otimes \mathbb R^{d_C}$. The corresponding theorem, which is fully proven in Appendix~\ref{appendix:classical}, can then be stated as follows:
\begin{theorem}\label{thm:ccec}
	Let $\vec p,\vec p' \in \mathbb R^d$ be two probability vectors with Shannon entropies $H(\vec p)$ and $H(\vec p')$, respectively. The following are equivalent:
	\begin{enumerate}
	\item For all $\epsilon>0$ we have $\vec p \rightarrow_\epsilon \vec p'$.
	\item $H(\vec p)\leq H(\vec p')$. 
	\end{enumerate}
\end{theorem}
Let me briefly comment on the main difference in the proof as compared to the quantum case: The essential construction of the catalyst is quite similar to the quantum case, however clearly we cannot make use of the Schur-Horn theorem, since we do not have access to unitary operations. 
The proof therefore proceeds by building into the catalyst a source of randomness, which instead of being used to dephase the system, is already correlated with the first part of the catalyst $S_2\cdots S_n A$ from the beginning and can be used to implement the transition $\vec p^{\otimes n} \rightarrow_{\varepsilon=0} \vec p'_{\epsilon,n}$ by a random permutation in the case that the auxiliary system $A$ is in state $n$. %(note that the epsilons in $\rightarrow_{\varepsilon=0}$ and $\vec p'_{\epsilon,n}$ are different quantities here). 
This source of randomness in general needs to have a dimension of the order of $d^n$ in contrast to the the quantum case, which only requires a dimension of the order of $\sqrt{d}$. 

\emph{Conclusion and open problems.} 
We have seen that von Neumann entropy and Shannon uniquely characterize state transitions that allow for the use of a catalyst but are otherwise reversible.
There are several natural open problems left for future work:
First, in Ref.~\cite{Boes2018b}, an exact form of the quantum catalytic entropy conjecture was conjectured, where no error is allowed in the transition from $\rho$ to $\rho'$ at the expense of the additional constraint that the rank of $\rho'$ has to be at least as large as that of $\rho$. 
This form of the conjecture seems much more difficult to prove and probably requires methods that go beyond standard typicality results (which always yield asymptotic statements with vanishingly small, but finite error).

A second open problem is to investigate whether a similar result also holds for other standard entropic quantities, such as mutual information or relative entropy. 

Finally, it will be worthwhile to explore the consequences of the given results for applications. Some {further} immediate applications of an affirmative resolution of the catalytic entropy conjecture in the context of (quantum) thermodynamics have been explored in Refs.~\cite{Boes2018b,Boes2020a}, but more applications can be expected. In particular, it would be interesting to see whether there are useful applications in the context of (potentially entanglement-assisted) communication scenarios.

\begin{acknowledgments} 
I would like to thank Paul Boes, Rodrigo Gallego, Markus P. M\"uller and Ivan Sergeev for extensive discussions about the catalytic entropy conjecture. In particular, I would like to thank Paul Boes for useful comments on an earlier draft and for suggesting
how the proof of the quantum result can be transferred to the classical setting.
I would further like to thank Joe Renes for pointing me to Ref.~\cite{Holenstein2011}.
This research was supported by the Swiss National Science Foundation through the National Centre of Competence in Research \emph{Quantum Science and Technology} (QSIT).

\emph{Note added: } After the publication of the first preprint of this work, Refs.~\cite{Kondra2021,LipkaBartosik2021} appeared that show similar single-shot characterizations of standard entropic quantities in the context of quantum teleportation and quantum entanglement.
\end{acknowledgments}
\bibliographystyle{apsrev4-1}

\begin{thebibliography}{48}%
\makeatletter
\providecommand \@ifxundefined [1]{%
 \@ifx{#1\undefined}
}%
\providecommand \@ifnum [1]{%
 \ifnum #1\expandafter \@firstoftwo
 \else \expandafter \@secondoftwo
 \fi
}%
\providecommand \@ifx [1]{%
 \ifx #1\expandafter \@firstoftwo
 \else \expandafter \@secondoftwo
 \fi
}%
\providecommand \natexlab [1]{#1}%
\providecommand \enquote  [1]{``#1''}%
\providecommand \bibnamefont  [1]{#1}%
\providecommand \bibfnamefont [1]{#1}%
\providecommand \citenamefont [1]{#1}%
\providecommand \href@noop [0]{\@secondoftwo}%
\providecommand \href [0]{\begingroup \@sanitize@url \@href}%
\providecommand \@href[1]{\@@startlink{#1}\@@href}%
\providecommand \@@href[1]{\endgroup#1\@@endlink}%
\providecommand \@sanitize@url [0]{\catcode `\\12\catcode `\$12\catcode
  `\&12\catcode `\#12\catcode `\^12\catcode `\_12\catcode `\%12\relax}%
\providecommand \@@startlink[1]{}%
\providecommand \@@endlink[0]{}%
\providecommand \url  [0]{\begingroup\@sanitize@url \@url }%
\providecommand \@url [1]{\endgroup\@href {#1}{\urlprefix }}%
\providecommand \urlprefix  [0]{URL }%
\providecommand \Eprint [0]{\href }%
\providecommand \doibase [0]{http://dx.doi.org/}%
\providecommand \selectlanguage [0]{\@gobble}%
\providecommand \bibinfo  [0]{\@secondoftwo}%
\providecommand \bibfield  [0]{\@secondoftwo}%
\providecommand \translation [1]{[#1]}%
\providecommand \BibitemOpen [0]{}%
\providecommand \bibitemStop [0]{}%
\providecommand \bibitemNoStop [0]{.\EOS\space}%
\providecommand \EOS [0]{\spacefactor3000\relax}%
\providecommand \BibitemShut  [1]{\csname bibitem#1\endcsname}%
\let\auto@bib@innerbib\@empty
%</preamble>
\bibitem [{\citenamefont {Bennett}\ \emph
  {et~al.}(1996{\natexlab{a}})\citenamefont {Bennett}, \citenamefont
  {Bernstein}, \citenamefont {Popescu},\ and\ \citenamefont
  {Schumacher}}]{Bennett1996}%
  \BibitemOpen
  \bibfield  {author} {\bibinfo {author} {\bibfnamefont {C.~H.}\ \bibnamefont
  {Bennett}}, \bibinfo {author} {\bibfnamefont {H.~J.}\ \bibnamefont
  {Bernstein}}, \bibinfo {author} {\bibfnamefont {S.}~\bibnamefont {Popescu}},
  \ and\ \bibinfo {author} {\bibfnamefont {B.}~\bibnamefont {Schumacher}},\
  }\href {\doibase 10.1103/physreva.53.2046} {\bibfield  {journal} {\bibinfo
  {journal} {Phys. Rev. A}\ }\textbf {\bibinfo {volume} {53}},\ \bibinfo
  {pages} {2046} (\bibinfo {year} {1996}{\natexlab{a}})}\BibitemShut {NoStop}%
\bibitem [{\citenamefont {Bennett}\ \emph
  {et~al.}(1996{\natexlab{b}})\citenamefont {Bennett}, \citenamefont
  {DiVincenzo}, \citenamefont {Smolin},\ and\ \citenamefont
  {Wootters}}]{Bennett1996a}%
  \BibitemOpen
  \bibfield  {author} {\bibinfo {author} {\bibfnamefont {C.~H.}\ \bibnamefont
  {Bennett}}, \bibinfo {author} {\bibfnamefont {D.~P.}\ \bibnamefont
  {DiVincenzo}}, \bibinfo {author} {\bibfnamefont {J.~A.}\ \bibnamefont
  {Smolin}}, \ and\ \bibinfo {author} {\bibfnamefont {W.~K.}\ \bibnamefont
  {Wootters}},\ }\href {\doibase 10.1103/physreva.54.3824} {\bibfield
  {journal} {\bibinfo  {journal} {Physical Review A}\ }\textbf {\bibinfo
  {volume} {54}},\ \bibinfo {pages} {3824} (\bibinfo {year}
  {1996}{\natexlab{b}})}\BibitemShut {NoStop}%
\bibitem [{\citenamefont {Vedral}\ \emph {et~al.}(1997)\citenamefont {Vedral},
  \citenamefont {Plenio}, \citenamefont {Rippin},\ and\ \citenamefont
  {Knight}}]{Vedral1997}%
  \BibitemOpen
  \bibfield  {author} {\bibinfo {author} {\bibfnamefont {V.}~\bibnamefont
  {Vedral}}, \bibinfo {author} {\bibfnamefont {M.~B.}\ \bibnamefont {Plenio}},
  \bibinfo {author} {\bibfnamefont {M.~A.}\ \bibnamefont {Rippin}}, \ and\
  \bibinfo {author} {\bibfnamefont {P.~L.}\ \bibnamefont {Knight}},\ }\href
  {\doibase 10.1103/physrevlett.78.2275} {\bibfield  {journal} {\bibinfo
  {journal} {Physical Review Letters}\ }\textbf {\bibinfo {volume} {78}},\
  \bibinfo {pages} {2275} (\bibinfo {year} {1997})}\BibitemShut {NoStop}%
\bibitem [{\citenamefont {Horodecki}\ \emph {et~al.}(2009)\citenamefont
  {Horodecki}, \citenamefont {Horodecki}, \citenamefont {Horodecki},\ and\
  \citenamefont {Horodecki}}]{Horodecki2009}%
  \BibitemOpen
  \bibfield  {author} {\bibinfo {author} {\bibfnamefont {R.}~\bibnamefont
  {Horodecki}}, \bibinfo {author} {\bibfnamefont {P.}~\bibnamefont
  {Horodecki}}, \bibinfo {author} {\bibfnamefont {M.}~\bibnamefont
  {Horodecki}}, \ and\ \bibinfo {author} {\bibfnamefont {K.}~\bibnamefont
  {Horodecki}},\ }\href {\doibase 10.1103/revmodphys.81.865} {\bibfield
  {journal} {\bibinfo  {journal} {Reviews of Modern Physics}\ }\textbf
  {\bibinfo {volume} {81}},\ \bibinfo {pages} {865} (\bibinfo {year}
  {2009})}\BibitemShut {NoStop}%
\bibitem [{\citenamefont {Lieb}\ and\ \citenamefont
  {Yngvason}(1999)}]{Lieb1999}%
  \BibitemOpen
  \bibfield  {author} {\bibinfo {author} {\bibfnamefont {E.~H.}\ \bibnamefont
  {Lieb}}\ and\ \bibinfo {author} {\bibfnamefont {J.}~\bibnamefont
  {Yngvason}},\ }\href {\doibase 10.1016/S0370-1573(98)00082-9} {\bibfield
  {journal} {\bibinfo  {journal} {Phys. Rep.}\ }\textbf {\bibinfo {volume}
  {310}},\ \bibinfo {pages} {1} (\bibinfo {year} {1999})}\BibitemShut {NoStop}%
\bibitem [{\citenamefont {Janzing}\ \emph {et~al.}(2000)\citenamefont
  {Janzing}, \citenamefont {Wocjan}, \citenamefont {Zeier}, \citenamefont
  {Geiss},\ and\ \citenamefont {Beth}}]{Janzing2000}%
  \BibitemOpen
  \bibfield  {author} {\bibinfo {author} {\bibfnamefont {D.}~\bibnamefont
  {Janzing}}, \bibinfo {author} {\bibfnamefont {P.}~\bibnamefont {Wocjan}},
  \bibinfo {author} {\bibfnamefont {R.}~\bibnamefont {Zeier}}, \bibinfo
  {author} {\bibfnamefont {R.}~\bibnamefont {Geiss}}, \ and\ \bibinfo {author}
  {\bibfnamefont {T.}~\bibnamefont {Beth}},\ }\href {\doibase
  10.1023/A:1026422630734} {\bibfield  {journal} {\bibinfo  {journal} {Int. J.
  Th. Phys.}\ }\textbf {\bibinfo {volume} {39}},\ \bibinfo {pages} {2717}
  (\bibinfo {year} {2000})}\BibitemShut {NoStop}%
\bibitem [{\citenamefont {Horodecki}\ and\ \citenamefont
  {Oppenheim}(2013)}]{Horodecki2013}%
  \BibitemOpen
  \bibfield  {author} {\bibinfo {author} {\bibfnamefont {M.}~\bibnamefont
  {Horodecki}}\ and\ \bibinfo {author} {\bibfnamefont {J.}~\bibnamefont
  {Oppenheim}},\ }\href {\doibase 10.1038/ncomms3059} {\bibfield  {journal}
  {\bibinfo  {journal} {Nature Comm.}\ }\textbf {\bibinfo {volume} {4}},\
  \bibinfo {pages} {2059} (\bibinfo {year} {2013})}\BibitemShut {NoStop}%
\bibitem [{\citenamefont {Brandao}\ \emph {et~al.}(2013)\citenamefont
  {Brandao}, \citenamefont {Horodecki}, \citenamefont {Oppenheim},
  \citenamefont {Renes},\ and\ \citenamefont {Spekkens}}]{Brandao2013}%
  \BibitemOpen
  \bibfield  {author} {\bibinfo {author} {\bibfnamefont {F.~G. S.~L.}\
  \bibnamefont {Brandao}}, \bibinfo {author} {\bibfnamefont {M.}~\bibnamefont
  {Horodecki}}, \bibinfo {author} {\bibfnamefont {J.}~\bibnamefont
  {Oppenheim}}, \bibinfo {author} {\bibfnamefont {J.~M.}\ \bibnamefont
  {Renes}}, \ and\ \bibinfo {author} {\bibfnamefont {R.~W.}\ \bibnamefont
  {Spekkens}},\ }\href {\doibase 10.1103/PhysRevLett.111.250404} {\bibfield
  {journal} {\bibinfo  {journal} {Phys. Rev. Lett.}\ }\textbf {\bibinfo
  {volume} {111}},\ \bibinfo {pages} {250404} (\bibinfo {year}
  {2013})}\BibitemShut {NoStop}%
\bibitem [{\citenamefont {Brandao}\ \emph {et~al.}(2015)\citenamefont
  {Brandao}, \citenamefont {Horodecki}, \citenamefont {Ng}, \citenamefont
  {Oppenheim},\ and\ \citenamefont {Wehner}}]{Brandao2015}%
  \BibitemOpen
  \bibfield  {author} {\bibinfo {author} {\bibfnamefont {F.~G. S.~L.}\
  \bibnamefont {Brandao}}, \bibinfo {author} {\bibfnamefont {M.}~\bibnamefont
  {Horodecki}}, \bibinfo {author} {\bibfnamefont {N.~H.~Y.}\ \bibnamefont
  {Ng}}, \bibinfo {author} {\bibfnamefont {J.}~\bibnamefont {Oppenheim}}, \
  and\ \bibinfo {author} {\bibfnamefont {S.}~\bibnamefont {Wehner}},\ }\href
  {\doibase 10.1073/pnas.1411728112} {\bibfield  {journal} {\bibinfo  {journal}
  {PNAS}\ }\textbf {\bibinfo {volume} {112}},\ \bibinfo {pages} {3275}
  (\bibinfo {year} {2015})}\BibitemShut {NoStop}%
\bibitem [{\citenamefont {Lostaglio}(2019)}]{Lostaglio2019}%
  \BibitemOpen
  \bibfield  {author} {\bibinfo {author} {\bibfnamefont {M.}~\bibnamefont
  {Lostaglio}},\ }\href {\doibase 10.1088/1361-6633/ab46e5} {\bibfield
  {journal} {\bibinfo  {journal} {Reports on Progress in Physics}\ }\textbf
  {\bibinfo {volume} {82}},\ \bibinfo {pages} {114001} (\bibinfo {year}
  {2019})}\BibitemShut {NoStop}%
\bibitem [{\citenamefont {Bartlett}\ \emph {et~al.}(2007)\citenamefont
  {Bartlett}, \citenamefont {Rudolph},\ and\ \citenamefont
  {Spekkens}}]{Bartlett2007}%
  \BibitemOpen
  \bibfield  {author} {\bibinfo {author} {\bibfnamefont {S.~D.}\ \bibnamefont
  {Bartlett}}, \bibinfo {author} {\bibfnamefont {T.}~\bibnamefont {Rudolph}}, \
  and\ \bibinfo {author} {\bibfnamefont {R.~W.}\ \bibnamefont {Spekkens}},\
  }\href {\doibase 10.1103/RevModPhys.79.555} {\bibfield  {journal} {\bibinfo
  {journal} {Rev. Mod. Phys.}\ }\textbf {\bibinfo {volume} {79}},\ \bibinfo
  {pages} {555} (\bibinfo {year} {2007})}\BibitemShut {NoStop}%
\bibitem [{\citenamefont {Marvian}\ and\ \citenamefont
  {Spekkens}(2014{\natexlab{a}})}]{Marvian2014}%
  \BibitemOpen
  \bibfield  {author} {\bibinfo {author} {\bibfnamefont {I.}~\bibnamefont
  {Marvian}}\ and\ \bibinfo {author} {\bibfnamefont {R.~W.}\ \bibnamefont
  {Spekkens}},\ }\href {\doibase 10.1103/PhysRevA.90.062110} {\bibfield
  {journal} {\bibinfo  {journal} {Phys. Rev. A}\ }\textbf {\bibinfo {volume}
  {90}},\ \bibinfo {pages} {062110} (\bibinfo {year}
  {2014}{\natexlab{a}})}\BibitemShut {NoStop}%
\bibitem [{\citenamefont {Marvian}\ and\ \citenamefont
  {Spekkens}(2014{\natexlab{b}})}]{Marvian2014a}%
  \BibitemOpen
  \bibfield  {author} {\bibinfo {author} {\bibfnamefont {I.}~\bibnamefont
  {Marvian}}\ and\ \bibinfo {author} {\bibfnamefont {R.~W.}\ \bibnamefont
  {Spekkens}},\ }\href {\doibase 10.1038/ncomms4821} {\bibfield  {journal}
  {\bibinfo  {journal} {Nature Comm.}\ }\textbf {\bibinfo {volume} {5}},\
  \bibinfo {pages} {3821} (\bibinfo {year} {2014}{\natexlab{b}})}\BibitemShut
  {NoStop}%
\bibitem [{\citenamefont {Chitambar}\ and\ \citenamefont
  {Gour}(2019)}]{Chitambar2019}%
  \BibitemOpen
  \bibfield  {author} {\bibinfo {author} {\bibfnamefont {E.}~\bibnamefont
  {Chitambar}}\ and\ \bibinfo {author} {\bibfnamefont {G.}~\bibnamefont
  {Gour}},\ }\href {\doibase 10.1103/revmodphys.91.025001} {\bibfield
  {journal} {\bibinfo  {journal} {Reviews of Modern Physics}\ }\textbf
  {\bibinfo {volume} {91}},\ \bibinfo {pages} {025001} (\bibinfo {year}
  {2019})}\BibitemShut {NoStop}%
\bibitem [{\citenamefont {Jonathan}\ and\ \citenamefont
  {Plenio}(1999)}]{Jonathan1999}%
  \BibitemOpen
  \bibfield  {author} {\bibinfo {author} {\bibfnamefont {D.}~\bibnamefont
  {Jonathan}}\ and\ \bibinfo {author} {\bibfnamefont {M.~B.}\ \bibnamefont
  {Plenio}},\ }\href {\doibase 10.1103/physrevlett.83.3566} {\bibfield
  {journal} {\bibinfo  {journal} {Physical Review Letters}\ }\textbf {\bibinfo
  {volume} {83}},\ \bibinfo {pages} {3566} (\bibinfo {year}
  {1999})}\BibitemShut {NoStop}%
\bibitem [{\citenamefont {Daftuar}\ and\ \citenamefont
  {Klimesh}(2001)}]{Daftuar2001}%
  \BibitemOpen
  \bibfield  {author} {\bibinfo {author} {\bibfnamefont {S.}~\bibnamefont
  {Daftuar}}\ and\ \bibinfo {author} {\bibfnamefont {M.}~\bibnamefont
  {Klimesh}},\ }\href {\doibase 10.1103/physreva.64.042314} {\bibfield
  {journal} {\bibinfo  {journal} {Physical Review A}\ }\textbf {\bibinfo
  {volume} {64}},\ \bibinfo {pages} {042314} (\bibinfo {year}
  {2001})}\BibitemShut {NoStop}%
\bibitem [{\citenamefont {van Dam}\ and\ \citenamefont
  {Hayden}(2003)}]{Dam2003}%
  \BibitemOpen
  \bibfield  {author} {\bibinfo {author} {\bibfnamefont {W.}~\bibnamefont {van
  Dam}}\ and\ \bibinfo {author} {\bibfnamefont {P.}~\bibnamefont {Hayden}},\
  }\href {\doibase 10.1103/physreva.67.060302} {\bibfield  {journal} {\bibinfo
  {journal} {Physical Review A}\ }\textbf {\bibinfo {volume} {67}},\ \bibinfo
  {pages} {060302} (\bibinfo {year} {2003})}\BibitemShut {NoStop}%
\bibitem [{\citenamefont {Aubrun}\ and\ \citenamefont
  {Nechita}(2007)}]{Aubrun2007}%
  \BibitemOpen
  \bibfield  {author} {\bibinfo {author} {\bibfnamefont {G.}~\bibnamefont
  {Aubrun}}\ and\ \bibinfo {author} {\bibfnamefont {I.}~\bibnamefont
  {Nechita}},\ }\href {\doibase 10.1007/s00220-007-0382-4} {\bibfield
  {journal} {\bibinfo  {journal} {Communications in Mathematical Physics}\
  }\textbf {\bibinfo {volume} {278}},\ \bibinfo {pages} {133} (\bibinfo {year}
  {2007})}\BibitemShut {NoStop}%
\bibitem [{\citenamefont {Klimesh}(2007)}]{Klimesh2007}%
  \BibitemOpen
  \bibfield  {author} {\bibinfo {author} {\bibfnamefont {M.}~\bibnamefont
  {Klimesh}},\ }\href@noop {} {\enquote {\bibinfo {title} {Inequalities that
  collectively completely characterize the catalytic majorization relation},}\
  } (\bibinfo {year} {2007}),\ \Eprint {http://arxiv.org/abs/0709.3680v1}
  {arXiv:0709.3680v1} \BibitemShut {NoStop}%
\bibitem [{\citenamefont {Turgut}(2007)}]{Turgut2007}%
  \BibitemOpen
  \bibfield  {author} {\bibinfo {author} {\bibfnamefont {S.}~\bibnamefont
  {Turgut}},\ }\href {\doibase 10.1088/1751-8113/40/40/012} {\bibfield
  {journal} {\bibinfo  {journal} {J. Phys. A}\ }\textbf {\bibinfo {volume}
  {40}},\ \bibinfo {pages} {12185} (\bibinfo {year} {2007})}\BibitemShut
  {NoStop}%
\bibitem [{\citenamefont {Ng}\ \emph {et~al.}(2015)\citenamefont {Ng},
  \citenamefont {Man\u{c}inska}, \citenamefont {Cirstoiu}, \citenamefont
  {Eisert},\ and\ \citenamefont {Wehner}}]{Ng2015}%
  \BibitemOpen
  \bibfield  {author} {\bibinfo {author} {\bibfnamefont {N.~H.~Y.}\
  \bibnamefont {Ng}}, \bibinfo {author} {\bibfnamefont {L.}~\bibnamefont
  {Man\u{c}inska}}, \bibinfo {author} {\bibfnamefont {C.}~\bibnamefont
  {Cirstoiu}}, \bibinfo {author} {\bibfnamefont {J.}~\bibnamefont {Eisert}}, \
  and\ \bibinfo {author} {\bibfnamefont {S.}~\bibnamefont {Wehner}},\ }\href
  {\doibase 10.1088/1367-2630/17/8/085004} {\bibfield  {journal} {\bibinfo
  {journal} {New J. Phys.}\ }\textbf {\bibinfo {volume} {17}},\ \bibinfo
  {pages} {085004} (\bibinfo {year} {2015})}\BibitemShut {NoStop}%
\bibitem [{\citenamefont {Boes}\ \emph {et~al.}(2019)\citenamefont {Boes},
  \citenamefont {Eisert}, \citenamefont {Gallego}, \citenamefont {Müller},\
  and\ \citenamefont {Wilming}}]{Boes2018b}%
  \BibitemOpen
  \bibfield  {author} {\bibinfo {author} {\bibfnamefont {P.}~\bibnamefont
  {Boes}}, \bibinfo {author} {\bibfnamefont {J.}~\bibnamefont {Eisert}},
  \bibinfo {author} {\bibfnamefont {R.}~\bibnamefont {Gallego}}, \bibinfo
  {author} {\bibfnamefont {M.~P.}\ \bibnamefont {Müller}}, \ and\ \bibinfo
  {author} {\bibfnamefont {H.}~\bibnamefont {Wilming}},\ }\href {\doibase
  10.1103/physrevlett.122.210402} {\bibfield  {journal} {\bibinfo  {journal}
  {Physical Review Letters}\ }\textbf {\bibinfo {volume} {122}},\ \bibinfo
  {pages} {210402} (\bibinfo {year} {2019})}\BibitemShut {NoStop}%
\bibitem [{\citenamefont {Sparaciari}\ \emph {et~al.}(2017)\citenamefont
  {Sparaciari}, \citenamefont {Jennings},\ and\ \citenamefont
  {Oppenheim}}]{Sparaciari2017}%
  \BibitemOpen
  \bibfield  {author} {\bibinfo {author} {\bibfnamefont {C.}~\bibnamefont
  {Sparaciari}}, \bibinfo {author} {\bibfnamefont {D.}~\bibnamefont
  {Jennings}}, \ and\ \bibinfo {author} {\bibfnamefont {J.}~\bibnamefont
  {Oppenheim}},\ }\href {\doibase 10.1038/s41467-017-01505-4} {\bibfield
  {journal} {\bibinfo  {journal} {Nature Communications}\ }\textbf {\bibinfo
  {volume} {8}} (\bibinfo {year} {2017}),\
  10.1038/s41467-017-01505-4}\BibitemShut {NoStop}%
\bibitem [{\citenamefont {Boes}\ \emph {et~al.}(2020)\citenamefont {Boes},
  \citenamefont {Gallego}, \citenamefont {Ng}, \citenamefont {Eisert},\ and\
  \citenamefont {Wilming}}]{Boes2020a}%
  \BibitemOpen
  \bibfield  {author} {\bibinfo {author} {\bibfnamefont {P.}~\bibnamefont
  {Boes}}, \bibinfo {author} {\bibfnamefont {R.}~\bibnamefont {Gallego}},
  \bibinfo {author} {\bibfnamefont {N.~H.~Y.}\ \bibnamefont {Ng}}, \bibinfo
  {author} {\bibfnamefont {J.}~\bibnamefont {Eisert}}, \ and\ \bibinfo {author}
  {\bibfnamefont {H.}~\bibnamefont {Wilming}},\ }\href {\doibase
  10.22331/q-2020-02-20-231} {\bibfield  {journal} {\bibinfo  {journal}
  {Quantum}\ }\textbf {\bibinfo {volume} {4}},\ \bibinfo {pages} {231}
  (\bibinfo {year} {2020})}\BibitemShut {NoStop}%
\bibitem [{\citenamefont {Shannon}(1948)}]{Shannon1948}%
  \BibitemOpen
  \bibfield  {author} {\bibinfo {author} {\bibfnamefont {C.~E.}\ \bibnamefont
  {Shannon}},\ }\href@noop {} {\bibfield  {journal} {\bibinfo  {journal} {The
  Bell System Technical Journal}\ }\textbf {\bibinfo {volume} {27}},\ \bibinfo
  {pages} {379} (\bibinfo {year} {1948})}\BibitemShut {NoStop}%
\bibitem [{\citenamefont {Thomas M.~Cover}(2006)}]{Cover2006}%
  \BibitemOpen
  \bibfield  {author} {\bibinfo {author} {\bibfnamefont {J.~A.~T.}\
  \bibnamefont {Thomas M.~Cover}},\ }\href
  {https://www.ebook.de/de/product/4435631/thomas_m_cover_joy_a_thomas_elements_of_information_theory.html}
  {\emph {\bibinfo {title} {Elements of Information Theory}}}\ (\bibinfo
  {publisher} {Wiley John + Sons},\ \bibinfo {year} {2006})\BibitemShut
  {NoStop}%
\bibitem [{\citenamefont {Wilde}(2009)}]{Wilde2009}%
  \BibitemOpen
  \bibfield  {author} {\bibinfo {author} {\bibfnamefont {M.~M.}\ \bibnamefont
  {Wilde}},\ }\href {\doibase 10.1017/cbo9781139525343} {\emph {\bibinfo
  {title} {Quantum Information Theory}}}\ (\bibinfo  {publisher} {Cambridge
  University Press},\ \bibinfo {year} {2009})\BibitemShut {NoStop}%
\bibitem [{\citenamefont {Gallego}\ \emph {et~al.}(2016)\citenamefont
  {Gallego}, \citenamefont {Eisert},\ and\ \citenamefont
  {Wilming}}]{Gallego2015}%
  \BibitemOpen
  \bibfield  {author} {\bibinfo {author} {\bibfnamefont {R.}~\bibnamefont
  {Gallego}}, \bibinfo {author} {\bibfnamefont {J.}~\bibnamefont {Eisert}}, \
  and\ \bibinfo {author} {\bibfnamefont {H.}~\bibnamefont {Wilming}},\ }\href
  {\doibase 10.1088/1367-2630/18/10/103017} {\bibfield  {journal} {\bibinfo
  {journal} {New J. Phys.}\ }\textbf {\bibinfo {volume} {18}},\ \bibinfo
  {pages} {103017} (\bibinfo {year} {2016})}\BibitemShut {NoStop}%
\bibitem [{\citenamefont {Wilming}\ \emph {et~al.}(2017)\citenamefont
  {Wilming}, \citenamefont {Gallego},\ and\ \citenamefont
  {Eisert}}]{Wilming2017a}%
  \BibitemOpen
  \bibfield  {author} {\bibinfo {author} {\bibfnamefont {H.}~\bibnamefont
  {Wilming}}, \bibinfo {author} {\bibfnamefont {R.}~\bibnamefont {Gallego}}, \
  and\ \bibinfo {author} {\bibfnamefont {J.}~\bibnamefont {Eisert}},\ }\href
  {\doibase 10.3390/e19060241} {\bibfield  {journal} {\bibinfo  {journal}
  {Entropy}\ }\textbf {\bibinfo {volume} {19}},\ \bibinfo {pages} {241}
  (\bibinfo {year} {2017})}\BibitemShut {NoStop}%
\bibitem [{\citenamefont {Lostaglio}\ \emph {et~al.}(2015)\citenamefont
  {Lostaglio}, \citenamefont {M\"uller},\ and\ \citenamefont
  {Pastena}}]{Lostaglio2015b}%
  \BibitemOpen
  \bibfield  {author} {\bibinfo {author} {\bibfnamefont {M.}~\bibnamefont
  {Lostaglio}}, \bibinfo {author} {\bibfnamefont {M.~P.}\ \bibnamefont
  {M\"uller}}, \ and\ \bibinfo {author} {\bibfnamefont {M.}~\bibnamefont
  {Pastena}},\ }\href {\doibase 10.1103/PhysRevLett.115.150402} {\bibfield
  {journal} {\bibinfo  {journal} {Phys. Rev. Lett.}\ }\textbf {\bibinfo
  {volume} {115}},\ \bibinfo {pages} {150402} (\bibinfo {year}
  {2015})}\BibitemShut {NoStop}%
\bibitem [{\citenamefont {M\"uller}\ and\ \citenamefont
  {Pastena}(2016)}]{Mueller2016}%
  \BibitemOpen
  \bibfield  {author} {\bibinfo {author} {\bibfnamefont {M.~P.}\ \bibnamefont
  {M\"uller}}\ and\ \bibinfo {author} {\bibfnamefont {M.}~\bibnamefont
  {Pastena}},\ }\href {\doibase 10.1109/tit.2016.2528285} {\bibfield  {journal}
  {\bibinfo  {journal} {{IEEE} Trans. Inf. Th.}\ }\textbf {\bibinfo {volume}
  {62}},\ \bibinfo {pages} {1711} (\bibinfo {year} {2016})}\BibitemShut
  {NoStop}%
\bibitem [{\citenamefont {Müller}(2018)}]{Mueller2017}%
  \BibitemOpen
  \bibfield  {author} {\bibinfo {author} {\bibfnamefont {M.~P.}\ \bibnamefont
  {Müller}},\ }\href {\doibase 10.1103/physrevx.8.041051} {\bibfield
  {journal} {\bibinfo  {journal} {Physical Review X}\ }\textbf {\bibinfo
  {volume} {8}},\ \bibinfo {pages} {041051} (\bibinfo {year}
  {2018})}\BibitemShut {NoStop}%
\bibitem [{\citenamefont {Rethinasamy}\ and\ \citenamefont
  {Wilde}(2020)}]{Rethinasamy2020}%
  \BibitemOpen
  \bibfield  {author} {\bibinfo {author} {\bibfnamefont {S.}~\bibnamefont
  {Rethinasamy}}\ and\ \bibinfo {author} {\bibfnamefont {M.~M.}\ \bibnamefont
  {Wilde}},\ }\href {\doibase 10.1103/physrevresearch.2.033455} {\bibfield
  {journal} {\bibinfo  {journal} {Physical Review Research}\ }\textbf {\bibinfo
  {volume} {2}},\ \bibinfo {pages} {033455} (\bibinfo {year}
  {2020})}\BibitemShut {NoStop}%
\bibitem [{\citenamefont {Shiraishi}\ and\ \citenamefont
  {Sagawa}(2021)}]{Shiraishi2020}%
  \BibitemOpen
  \bibfield  {author} {\bibinfo {author} {\bibfnamefont {N.}~\bibnamefont
  {Shiraishi}}\ and\ \bibinfo {author} {\bibfnamefont {T.}~\bibnamefont
  {Sagawa}},\ }\href {\doibase 10.1103/physrevlett.126.150502} {\bibfield
  {journal} {\bibinfo  {journal} {Physical Review Letters}\ }\textbf {\bibinfo
  {volume} {126}},\ \bibinfo {pages} {150502} (\bibinfo {year}
  {2021})}\BibitemShut {NoStop}%
%\bibitem [{Sup()}]{SuppMat}%
%  \BibitemOpen
%  \href@noop {} {}\bibinfo {note} {See Supplemental Material at [url will be
%  inserted] for a proof sketch of Lemma 4, a detailed discussion of how to
%  construct the unitary $U$, the formulation and proof of the classical version
%  of the main result, a discussion of the dimension required for the catalyst
%  and Refs.~\cite{Boes2020,Marshall2011a}.}\BibitemShut {Stop}%
\bibitem [{\citenamefont {Fannes}(1973)}]{Fannes1973}%
  \BibitemOpen
  \bibfield  {author} {\bibinfo {author} {\bibfnamefont {M.}~\bibnamefont
  {Fannes}},\ }\href {\doibase 10.1007/bf01646490} {\bibfield  {journal}
  {\bibinfo  {journal} {Communications in Mathematical Physics}\ }\textbf
  {\bibinfo {volume} {31}},\ \bibinfo {pages} {291} (\bibinfo {year}
  {1973})}\BibitemShut {NoStop}%
\bibitem [{\citenamefont {Audenaert}(2007)}]{Audenaert2007}%
  \BibitemOpen
  \bibfield  {author} {\bibinfo {author} {\bibfnamefont {K.~M.~R.}\
  \bibnamefont {Audenaert}},\ }\href {\doibase 10.1088/1751-8113/40/28/s18}
  {\bibfield  {journal} {\bibinfo  {journal} {Journal of Physics A:
  Mathematical and Theoretical}\ }\textbf {\bibinfo {volume} {40}},\ \bibinfo
  {pages} {8127} (\bibinfo {year} {2007})}\BibitemShut {NoStop}%
\bibitem [{\citenamefont {Kammerlander}\ and\ \citenamefont
  {Renner}(2020)}]{Kammerlander2020}%
  \BibitemOpen
  \bibfield  {author} {\bibinfo {author} {\bibfnamefont {P.}~\bibnamefont
  {Kammerlander}}\ and\ \bibinfo {author} {\bibfnamefont {R.}~\bibnamefont
  {Renner}},\ }\href@noop {} {\  (\bibinfo {year} {2020})},\ \Eprint
  {http://arxiv.org/abs/2002.08968} {arXiv:2002.08968 [math-ph]} \BibitemShut
  {NoStop}%
\bibitem [{\citenamefont {Alicki}\ and\ \citenamefont
  {Fannes}(2013)}]{Alicki2013}%
  \BibitemOpen
  \bibfield  {author} {\bibinfo {author} {\bibfnamefont {R.}~\bibnamefont
  {Alicki}}\ and\ \bibinfo {author} {\bibfnamefont {M.}~\bibnamefont
  {Fannes}},\ }\href {\doibase 10.1103/physreve.87.042123} {\bibfield
  {journal} {\bibinfo  {journal} {Physical Review E}\ }\textbf {\bibinfo
  {volume} {87}},\ \bibinfo {pages} {042123} (\bibinfo {year}
  {2013})}\BibitemShut {NoStop}%
\bibitem [{\citenamefont {Pusz}\ and\ \citenamefont
  {Woronowicz}(1978)}]{Pusz1978}%
  \BibitemOpen
  \bibfield  {author} {\bibinfo {author} {\bibfnamefont {W.}~\bibnamefont
  {Pusz}}\ and\ \bibinfo {author} {\bibfnamefont {S.}~\bibnamefont
  {Woronowicz}},\ }\href {\doibase 10.1007/BF01614224} {\bibfield  {journal}
  {\bibinfo  {journal} {Commun. Math. Phys.}\ }\textbf {\bibinfo {volume}
  {58}},\ \bibinfo {pages} {273} (\bibinfo {year} {1978})}\BibitemShut
  {NoStop}%
\bibitem [{\citenamefont {Lenard}(1978)}]{Lenard1978}%
  \BibitemOpen
  \bibfield  {author} {\bibinfo {author} {\bibfnamefont {A.}~\bibnamefont
  {Lenard}},\ }\href {\doibase 10.1007/BF01011769} {\bibfield  {journal}
  {\bibinfo  {journal} {J. Stat. Phys.}\ }\textbf {\bibinfo {volume} {19}},\
  \bibinfo {pages} {575} (\bibinfo {year} {1978})}\BibitemShut {NoStop}%
\bibitem [{foo()}]{footnotepassive}%
  \BibitemOpen
  \href@noop {} {}\bibinfo {note} {In \cite{Sparaciari2017} it was argued that
  by concatenating arbitrarily many catalytic transitions the bound in
  Corollary~\ref{cor:passive} can be achieved. However, concatenating two
  catalytic transitions builds up correlations between the catalysts and thus
  is in general not catalytic as a whole. Our main result shows that one can
  nevertheless always find a catalytic transformation implementing the two
  steps in one.}\BibitemShut {Stop}%
\bibitem [{\citenamefont {Holenstein}\ and\ \citenamefont
  {Renner}(2011)}]{Holenstein2011}%
  \BibitemOpen
  \bibfield  {author} {\bibinfo {author} {\bibfnamefont {T.}~\bibnamefont
  {Holenstein}}\ and\ \bibinfo {author} {\bibfnamefont {R.}~\bibnamefont
  {Renner}},\ }\href {\doibase 10.1109/tit.2011.2110230} {\bibfield  {journal}
  {\bibinfo  {journal} {{IEEE} Transactions on Information Theory}\ }\textbf
  {\bibinfo {volume} {57}},\ \bibinfo {pages} {1865} (\bibinfo {year}
  {2011})}\BibitemShut {NoStop}%
\bibitem [{\citenamefont {Boes}\ \emph {et~al.}(2018)\citenamefont {Boes},
  \citenamefont {Wilming}, \citenamefont {Gallego},\ and\ \citenamefont
  {Eisert}}]{Boes2018a}%
  \BibitemOpen
  \bibfield  {author} {\bibinfo {author} {\bibfnamefont {P.}~\bibnamefont
  {Boes}}, \bibinfo {author} {\bibfnamefont {H.}~\bibnamefont {Wilming}},
  \bibinfo {author} {\bibfnamefont {R.}~\bibnamefont {Gallego}}, \ and\
  \bibinfo {author} {\bibfnamefont {J.}~\bibnamefont {Eisert}},\ }\href
  {\doibase 10.1103/physrevx.8.041016} {\bibfield  {journal} {\bibinfo
  {journal} {Physical Review X}\ }\textbf {\bibinfo {volume} {8}},\ \bibinfo
  {pages} {041016} (\bibinfo {year} {2018})}\BibitemShut {NoStop}%
\bibitem [{\citenamefont {Kondra}\ \emph {et~al.}(2021)\citenamefont {Kondra},
  \citenamefont {Datta},\ and\ \citenamefont {Streltsov}}]{Kondra2021}%
  \BibitemOpen
  \bibfield  {author} {\bibinfo {author} {\bibfnamefont {T.~V.}\ \bibnamefont
  {Kondra}}, \bibinfo {author} {\bibfnamefont {C.}~\bibnamefont {Datta}}, \
  and\ \bibinfo {author} {\bibfnamefont {A.}~\bibnamefont {Streltsov}},\
  }\href {\doibase 10.1103/PhysRevLett.127.150503} {\bibfield
  {journal} {\bibinfo  {journal} {Physical Review Letters}\ }\textbf {\bibinfo
  {volume} {127}},\ \bibinfo {pages} {150502} (\bibinfo {year}
  {2021})}\BibitemShut {NoStop}%
\bibitem [{\citenamefont {Lipka-Bartosik}\ and\ \citenamefont
  {Skrzypczyk}(2021)}]{LipkaBartosik2021}%
  \BibitemOpen
  \bibfield  {author} {\bibinfo {author} {\bibfnamefont {P.}~\bibnamefont
  {Lipka-Bartosik}}\ and\ \bibinfo {author} {\bibfnamefont {P.}~\bibnamefont
  {Skrzypczyk}},\ }\href {\doibase 10.1103/physrevlett.127.080502} {\bibfield
  {journal} {\bibinfo  {journal} {Physical Review Letters}\ }\textbf {\bibinfo
  {volume} {127}},\ \bibinfo {pages} {080502} (\bibinfo {year}
  {2021})}\BibitemShut {NoStop}%
\bibitem [{\citenamefont {Boes}\ \emph {et~al.}()\citenamefont {Boes},
  \citenamefont {Ng},\ and\ \citenamefont {Wilming}}]{Boes2020}%
  \BibitemOpen
  \bibfield  {author} {\bibinfo {author} {\bibfnamefont {P.}~\bibnamefont
  {Boes}}, \bibinfo {author} {\bibfnamefont {N.~H.~Y.}\ \bibnamefont {Ng}}, \
  and\ \bibinfo {author} {\bibfnamefont {H.}~\bibnamefont {Wilming}},\
  }\href@noop {} {\enquote {\bibinfo {title} {The variance of relative
  surprisal as single-shot quantifier},}\ }\Eprint
  {http://arxiv.org/abs/2009.08391} {arXiv:2009.08391} \BibitemShut {NoStop}%
\bibitem [{\citenamefont {Marshall}\ \emph {et~al.}(2011)\citenamefont
  {Marshall}, \citenamefont {Olkin},\ and\ \citenamefont
  {Arnold}}]{Marshall2011a}%
  \BibitemOpen
  \bibfield  {author} {\bibinfo {author} {\bibfnamefont {A.~W.}\ \bibnamefont
  {Marshall}}, \bibinfo {author} {\bibfnamefont {I.}~\bibnamefont {Olkin}}, \
  and\ \bibinfo {author} {\bibfnamefont {B.~C.}\ \bibnamefont {Arnold}},\
  }\href {\doibase 10.1007/978-0-387-68276-1} {\emph {\bibinfo {title}
  {Inequalities: Theory of Majorization and Its Applications}}}\ (\bibinfo
  {publisher} {Springer New York},\ \bibinfo {year} {2011})\BibitemShut
  {NoStop}%
\end{thebibliography}
\clearpage
\appendix

\section{Proof-sketch of Lemma~4}
\label{appendix:lemma4}
We here provide a sketch of the proof of Lemma~\ref{lemma:typmaj}. 
For any $\delta>0$ and state $\rho=\sum_i p_i \proj{i}$ define the typical subspace $\Pi_\delta^{\rho^{\otimes n}}$ as the subspace spanned by those states $\ket{i_1}\otimes\cdots\otimes \ket{i_n}$ such that 
\begin{align}
	\left|\frac{1}{n} \sum_{j=1}^n \log(1/p_{i_j})-H(\rho)\right| < \delta.
\end{align}
We further identify $\Pi_\delta^{\rho^{\otimes n}}$ with the projector onto the subspace.
It follows from Hoeffding's inequality that $\rho^{\otimes n}$ is approximated by 
\begin{align}
	\hat \rho_{\delta,n}:=\frac{\Pi_\delta^{\rho^{\otimes n}} \rho^{\otimes n} \Pi_\delta^{\rho^{\otimes n}}}{\Tr[\Pi_\delta^{\rho^\otimes n} \rho^{\otimes n}]}
\end{align}
with an error in trace-distance of the order $\exp(-n \delta^2)$ and this error is essentially optimal up to constants for general density matrices \cite{Holenstein2011} (but of course may deviate significantly for particular density matrices). 
Note that constructing $\hat \rho_{\delta,n}$ simply consists of the following steps:
\begin{enumerate}
	\item Take $\rho^{\otimes n}$ with eigenvalues of the form $q_{i_1\ldots i_n}=p_{i_1}\cdots p_{i_n}$ and write it out in its eigenbasis.
	\item Go through all diagonal elements $q_{i_1\ldots i_n}$ and if $$\left|-\frac{1}{n} \log(q_{i_1\ldots i_n}) - H(\rho)\right|\geq \delta$$ replace it with $0$. Otherwise keep it.
	\item Renormalize the density matrix.
\end{enumerate}

Furthermore, the following properties are well-known \cite{Wilde2009}:  i) For any $\varepsilon>0$ and sufficiently large $n$, we have
\begin{align}
	(1-\varepsilon) \e^{n(H(\rho)-\delta)}\leq \Tr[\Pi_\delta^{\rho^{\otimes n}}]\leq \e^{n(H(\rho)+\delta)}.
\end{align}
ii) We have the operator inequality
\begin{align}
	\e^{-n(H(\rho)+\delta)}\Pi_\delta^{\rho^{\otimes n}} \leq \Pi_\delta^{\rho^{\otimes n}} \rho^{\otimes n} \Pi_\delta^{\rho^{\otimes n}} \leq \e^{-n(H(\rho)-\delta)}\Pi_\delta^{\rho^{\otimes n}}.
\end{align}
iii) The approximation $\hat \rho_{\delta,n}$ is diagonal in the same basis as $\rho^{\otimes n}$, i.e. we can choose $\mc D_{\hat \rho_{\delta,n}}=\mc D_\rho^{\otimes n}$.

These conditions ensure that the approximation $\hat \rho_{\delta,n}$ majorizes the approximation $\hat \rho'_{\delta,n}$ if
\begin{align}
	\Delta H +\frac{\log(1-\epsilon)}{2n} > 2\delta. 
\end{align}
Thus, if we choose $\delta = \Delta H/4$ and $n$ large enough, $\hat \rho_{\delta,n} \succeq \hat \rho'_{\delta,n}$.
By the Schur-Horn theorem, there now exists a unitary $U$ such that
\begin{align}
	\hat\rho'_{\delta,n}=	\mc D_{\hat \rho'_{\delta,n}}[U \hat \rho_{\delta,n} U^\dagger] =  	\mc D_{\rho'}^{\otimes n}[U \hat \rho_{\delta,n} U^\dagger]=: \mathcal C[\hat\rho_{\delta,n}].
\end{align}
An explicit construction of $U$ given $\hat \rho_{\delta,n}$ and $\hat \rho'_{\delta,n}$ is provided in the next section.

We now set
\begin{align}
	\rho'_{\epsilon,n} := \mc C[\rho^{\otimes n}]=\mc D_{\rho'}^{\otimes n}[U \rho^{\otimes n} U^\dagger].
\end{align}
Then, since the trace-distance is non-increasing under completely positive, trace-preserving maps and using the triangle inequality, we get
\begin{align}
	D\big(\rho'_{\epsilon,n}, \rho'^{\otimes n}\big)&\leq D\big(\rho'_{\epsilon,n},\rho'_{\delta,n}\big) + D\big(\rho'_{\delta,n},\rho'^{\otimes n}\big)\\
	&= D\big(\mc C[\rho^{\otimes n}],\mc C[\rho_{\delta,n}]\big) +  D\big(\rho'_{\delta,n},\rho'^{\otimes n}\big) \nonumber\\
	&\leq D\big(\rho^{\otimes n},\rho_{\delta,n}\big) + D\big(\rho'_{\delta,n},\rho'^{\otimes n}\big) \\
	&\leq O(\exp(-n \delta^2)+  O(\exp(-n \delta^2)\\
	&=O(\exp(-n\Delta H^2/4).
\end{align}
Note that in particular we have $\rho'_{\epsilon,n} = \mc D_{\rho'}^{\otimes n}[\rho'_{\epsilon,n}]$ and that the unitary $U$ is the same unitary that is also used in the construction of the catalyst, i.e., for the state $\chi = U\rho^{\otimes n} U^\dagger$.

\section{How to explicitly construct unitaries implementing majorization and the complexity of our constructions}
\label{appendix:howto}
We now discuss in more detail the explicit construction of the relvant unitary operator and catalyst state for a given catalytic transition $\rho\rightarrow \rho'$.
Essentially without loss of generality we will assume that $\rho$ and $\rho'$ are diagonal in the same basis, since we can always adjust the final eigenbasis by a unitary without affecting the catalyst.

First, we remind the reader the construction of the catalyst and unitary given in the main text consists of two steps. 
In the first step, a catalyst and unitary is constructed whose effect is to bring the system to a state whose diagonal coincides with the final state. 
In the second step a second, uncorrelated part of the catalyst, which is maximally mixed, is used to decohere the system in the eigenbasis of the final state. 
The construction for the second step only depends on the eigenbasis of $\rho'$ and consists of dilating the decoherence-channel to a unitary operation. An explicit and optimal (in terms of Hilbert-space dimension) construction can be found in Ref.~\cite{Boes2018a} and I will omit discussing this second step in more detail. 

For the first step, the state of the catalyst is given by:
\begin{align}
	\sigma_1 &= \frac{1}{n}\sum_{k=1}^n \rho^{\otimes k-1}\otimes \chi_{1,\ldots,n-k}\otimes \proj{k}_A\\
	&=\frac{1}{n}\sum_{k=1}^n \rho^{\otimes k-1}\otimes \Tr_{n-k+1\ldots n}[U \rho^{\otimes n} U^\dagger]\otimes \proj{k}_A.
\end{align}
Note that the state $\sigma_1$ is constructed from three ingredients: the counting-register $A$, copies of the input state $\rho$ and (reduced states of) the state $\chi = U \rho^{\otimes n} U^\dagger$. Here, $U$ is the unitary that implements the majorization $\rho^{\otimes n}\succeq \rho'_{\epsilon,n}$ in the sense that $\chi$ coincides with $\rho'_{\epsilon,n}$ in the eigenbasis of ${\rho'}^{\otimes n}$. The construction of $\rho'_{\epsilon,n}$ is made explicit (up to the precise choice of $n$ as a function of $\epsilon$) in the previous section.
The final unitary that is applied to system and catalyst can be written as
\begin{align}
	W = P \left[U\otimes\proj{n}_A + \sum_{k=1}^{n-1}\mathbf 1\otimes \proj{k}\right],
\end{align}
where $P$ is the joint permutation matrix representing steps 2. and 3. in the main text. 

It therefore remains to discuss how to explicitly construct the unitary $U$ (which is the same as in the previous section). For the convenience of the reader, I now explain this construction for the general case of two density matrices $\omega \succeq \omega'$. To obtain $U$ one then applies this procedure to $\hat \rho_{\delta,n} \succeq \hat \rho'_{\delta,n}$ (for sufficiently large $n$). 
That is, given that $\omega \succeq \omega'$ we now look for the unitary $U$ such that $U\omega U^\dagger$ coincides on the diagonal with $\omega '$ (in the eigenbasis of $\omega'$).
For simplicity we assume again that $\omega$ and $\omega'$ are diagonal in the same basis $\{\ket i\}$ (otherwise we simply add a unitary rotation at the end), so that
\begin{align}
	\omega = \sum_i w_i \proj{i},\quad \omega' = \sum_i w_i' \proj{i}. 
\end{align}
We further assume that $w_1 \geq w_2 \geq \ldots$ and $w_1'\geq w_2' \geq \ldots$, which again can be ensured by a unitary transformation.
The condition $\omega\succeq \omega'$ now is equivalent to the condition $\vec w \succeq \vec w'$ as mentioned in the main text, where $\vec w = (w_1,\ldots,w_d)^\top$ and $\vec w'=(w_1',\ldots,w_d')^\top$ are probability vectors.
%This condition on the other hand is equivalent to
%\begin{align}
%	\sum_{j=1}^k w_j \geq \sum_{j=1}^k w_j'\quad \forall k=1,\ldots,d.
%\end{align}
%Indeed, the latter condition is often taken as the definition of majorization \cite{}.
We now need to introduce the notion of a \emph{T-transform} on a pair of indicies $(j,k)$ as the stochastic matrix acting as
\begin{align}
	(T_{(j,k)}(t) \vec w)_i &= 
		\begin{cases}
			w_i \quad &\text{if}\quad i\neq j,k\\
			t w_j + (1-t)w_k &\text{if}\quad i=j\\
			t w_k + (1-t)w_j &\text{if}\quad i=k.
		\end{cases}
\end{align}
That is, the T-transform transposes two entries of a probability vector with probability $(1-t)$ and does nothing to the remaing entries. Note that for $t=1$ the T-transform does nothing, while for $t=0$ it interchanges the two entries, and for $t=1/2$ it mixes them equally. 
Importantly, if $\vec w \succeq \vec w'$, then $\vec w'$ can be obtained by applying at most $d-1$ suitable T-transforms to the probability vector $\vec w$.
Concretely, we can use the following algorithm involving a loop over a counting variable $i$ which starts with $i=1$ \cite{Marshall2011a}:
\begin{enumerate}
	\item\label{item:startalgo} If $i=1$ set $\vec w^{(i)}:=\vec w$. 
	\item If $\vec w^{(i)}=\vec w'$ stop. Otherwise continue:
	\item Set $k^{(i)}$ to be the smallest integer such that $w'_{k^{(i)}} > w^{(i)}_{k^{(i)}}$ and $j^{(i)}$ the largest integer such that $w'_{j^{(i)}} < w^{(i)}_{j^{(i)}}$. 
	\item Set $\delta=\min{(w^{(i)}_{j^{(i)}} - w_{j^{(i)}}',w_{k^{(i)}}'-w^{(i)}_{k^{(i)}})}$ and $t^{(i)} =1- \delta/(w^{(i)}_{j^{(i)}}-w^{(i)}_{k^{(i)}})$.
	\item Set $\vec w^{(i+1)} := T_{(j^{(i)},k^{(i)})}(t^{(i)})\vec w^{(i)}$.
\item Increase $i\mapsto i+1$. Go to step \ref{item:startalgo}.
\end{enumerate}
It is shown in Ref.~\cite{Marshall2011a} that the algorithm terminates in at most $d-1$ steps. 
Collecting the different parameters $j^{(i)},k^{(i)}$ and $t^{(i)}$, we find
\begin{align}
	\vec w' = T_{(j^{(d-1)},k^{(d-1)})}(t^{(d-1)})\cdots T_{(j^{(1)},k^{(1)})}(t^{(1)})\vec w.
\end{align}
For the benefit of the reader, below I list Wolfram Mathematica code that performs a single step of the loop. I.e., it calculates the parameters $j,k,t$ to go one step from a vector $\vec w$ to a vector $\vec q$ and applies the corresponding T-transform to $\vec w$.
\lstset{frame=tb,
  language=Mathematica,
  aboveskip=3mm,
  belowskip=3mm,
  showstringspaces=false,
  columns=flexible,
  basicstyle={\small\ttfamily},
  numbers=none,
  numberstyle=\tiny\color{gray},
  keywordstyle=\color{blue},
  commentstyle=\it,
  stringstyle=\color{mauve},
  breaklines=true,
  breakatwhitespace=true,
  tabsize=3
}
\begin{lstlisting}
(* Helper Function: Returns position of last entry in list x for which pattern pat is true *)
LastPosition[x_, pat_] :=
   Length[x] + 1 - FirstPosition[Reverse[x], pat];

(* T-Transform applied to vector w*)
T[w_, {j_, k_, t_}] := 
   t w + (1 - t) Permute[w, Cycles[{{j, k}}]];

(* Computes values for {j,k,t} to make one step in majorization space from w to the direction of q *)
OneStep[w_, q_] := Module[{j, k, delta, t},
   If[w != q, 
    k = FirstPosition[q - w, _?Positive][[1]];
    j = LastPosition[w - q, _?Positive][[1]];
    delta = Min[{w[[j]] - q[[j]], q[[k]] - w[[k]]}];
    t = 1 - delta/(w[[j]] - w[[k]]);
    {j, k, t}
    , w]
   ];

(* Apply T-transform to go to obtain next vector in loop *)
T[w, OneStep[w, q]]
\end{lstlisting}

Importantly, the $T$ transform can be implemented unitarily as a rotation matrix on the subspace spanned by $\ket{j}$ and $\ket{k}$. I.e., if we consider the unitary matrix $U_{(j,k)}(t)$ that acts as
\begin{align}
		U_{(j,k)}(t) \ket{i} &= 
		\begin{cases}
			\ket i \quad &\text{if}\quad i\neq j,k\\
			\sqrt{t}\ket{j} + \sqrt{1-t}\ket{k} &\text{if}\quad i=j\\
			 \sqrt{t}\ket{k} - \sqrt{1-t}\ket{j} &\text{if}\quad i=k,
		\end{cases}
\end{align}
then the density matrix $U_{(j,k)}(t) \omega U_{(j,k)}(t)^\dagger$ has as diagonal entries the vector $T_{(j,k)}(t)\vec w$.
Correspondingly, we find that if we set
\begin{align}
	U = U_{(j^{(d-1)},k^{(d-1)})}(t^{(d-1)})\cdots U_{(j^{(1)},k^{(1)})}(t^{(1)}),
\end{align}
then the density matrix $U\omega U^\dagger$ has as as diagonal elements the vector
\begin{align}
T_{(j^{(d-1)},k^{(d-1)})}(t^{(d-1)})\cdots T_{(j^{(1)},k^{(1)})}(t^{(1)})\vec w = \vec w'.
\end{align}
as we wished for. This finishes showing how to explicitly construct the majorization unitary.

Let us finally comment on the difficulty of numerically constructing classical descriptions of all the objects in practice for a given transition $\rho\rightarrow \rho'$.
First it is clear that if the dimension $d$ of the Hilbert-space of $\rho$ is very large, then even finding the sepctrum of $\rho$, and hence computing the entropy of $\rho$ is computationally hard.
For example, if $\rho$ would be the thermal state of a quantum many-body system, specified by some local Hamiltonian, then computing either the spectrum of $\rho$ or the entropy of $\rho$ is typically completely intractable.
However, this is not specific to the problem at hand but true for any problem that involves the spectrum or von Neumann entropy of the density matrix of a many-body system (and hence many problems in quantum information theory).

Nevertheless, even when $d$ is not large, it is computationally difficult to compute the catalyst and unitary used in the proof in this paper. This is due to the fact that the construction of the catalyst involves a large number $n$ of copies of $\rho$ to construct the states $\hat \rho_{\delta,n}$ and $\hat\rho'_{\delta,n}$ from which $U$ is computed. In other words, the Hilbert-space dimension of the catalyst constructed here grows exponentially with $n$ (prohibiting to write down concrete matrix representations of all the involved objects) and taking large $n$ is necessary to achieve good precision $\epsilon$ in general.
It is important to realize, however, that the catalyst and unitaries constructed here will often not be the optimal ones.
To illustrate this, consider the initial state $\rho$ with eigenvalues $(1/2,1/2,0)$ and final state $\rho'$ with eigenvalues $(2/3,1/6,1/6)$,
both diagonal in the basis $\{\ket{1},\ket{2},\ket{3}\}$.
It is possible to implement the transition $\rho\rightarrow \rho'$ \emph{exactly} with a small catalyst of dimension $d_C=2$, namely $\sigma$ with eigenvalues $(2/3,1/3)$ and eigenbasis $\{\ket{1}_C,\ket{2}_C\}$.
The unitary $U$ that implements this transition acts as
\begin{align}
	U\ket{2}\otimes \ket{1}_C &= \ket 1\otimes \ket{2}_C,\\
	U\ket 1\otimes \ket{2}_C &= \ket 2\otimes \ket{1}_C,\\
	U\ket 3\otimes \ket{1}_C &= \ket 2\otimes \ket{2}_C,\\
	U \ket 2\otimes \ket{2}_C &= \ket 3\otimes \ket{1}_C
\end{align}
and $U\ket{i}\otimes\ket{j}_C=\ket{i}\otimes\ket{j}_C$ on the remaining basis vectors.
Nevertheless, the construction presented in this paper will only yield an approximate transition $\rho\mapsto \rho'_\epsilon$ with arbitrarily small $\epsilon$ provided that $n$ (and hence the catalyst dimension) is large.
It is therefore an interesting open problem to find \emph{optimal} constructions in terms of Hilbert-space dimension of catalysts for given transitions $\rho\rightarrow \rho'$. Indeed, it is currently unknown whether in the case $H(\rho)<H(\rho')$ one can achieve perfect conversion with $\epsilon=0$ with a finite-dimensional catalyst (see also Section~\ref{sec:size}).

\section{The classical result}
\label{appendix:classical}
In this section we formulate catalytic transitions in the classical setting and prove the corresponding characterization of Shannon entropy. 
We continue to speak of systems, and a system $S$ is described by a probability vector $\vec p \in \mathbb R^{|S|}$, where $|S|$ denotes the number of distinct states of $S$.
Joint states of two systems $S$ and $S'$ are described by probability vectors in $\mathbb R^{|S_1|}\otimes \mathbb R^{|S_2|}\simeq \mathbb R^{|S_1||S_2|}$.
If $\vec p_{SS'} \in\mathbb R^{|S|}\otimes\mathbb R^{|S'|}$ is a probability vector describing a bipartite state, we continue to write
\begin{align}
	\vec p_S := \tr_{S'}[\vec p_{SS'}] 
\end{align}
for the marginal on $S$, whose entries are given by
\begin{align}
	(\vec p_S)_i = \sum_{j=1}^{|S'|} (\vec p_{SS'})_{i,j}.
\end{align}
Instead of unitary matrices acting on $\mathbb C^d$, we now continue permutation matrices $\pi$ acting on $\mathbb R^d$ by re-shuffling the basis vectors:
\begin{align}
	(\pi \vec p)_i = \sum_{j=1}^d \pi_{i,j} p_j,
\end{align}
where $\pi$ is a permutation matrix and $\vec p=(p_1,\ldots,p_d)^\top$.
%%%%%%%%%%%%%%%%%%%%%%
\begin{figure*}[t!]
	\includegraphics[width=12cm]{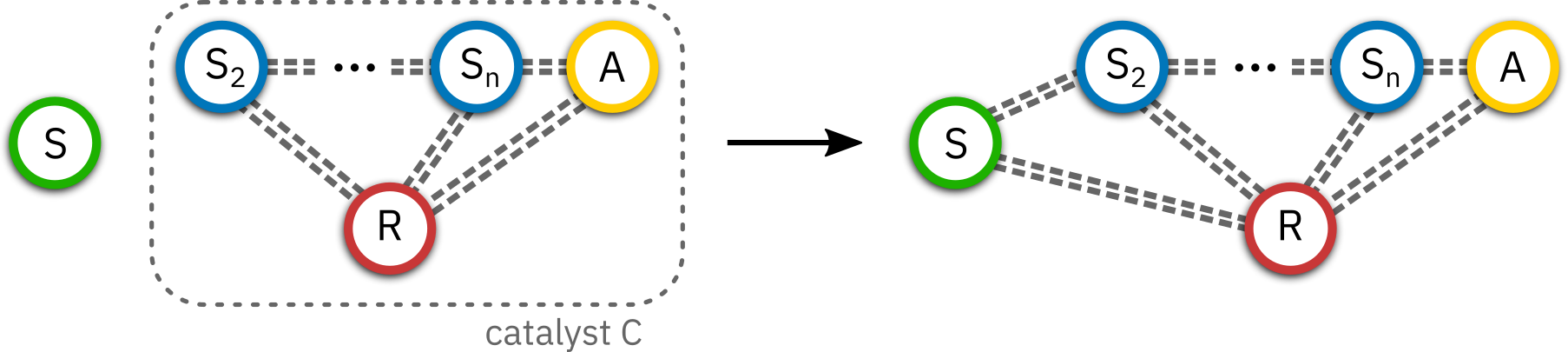}
	\caption{The structure of the constructed catalyst $C$ in the classical case: It contains subsystems $S_2,\ldots, S_n$, which are copies of the target system $S$ together with an auxiliary system $A$ and a catalytic source of randomness $R$. The dashed lines indicate possible correlations. The source of randomness is now utilized to fuel a random permutation on $S_1\cdots S_n$ to make use of majorization.  The catalyst and global permutation $\pi$ are constructed in such a way  that the correlations between all sub-systems of the catalyst remain invariant.}
		\label{fig:structure_catalyst_c}
\end{figure*}
%%%%%%%%%%%%%%%%%%%%%5
For two vectors $\vec p,\vec p'\in \mathbb R^d$ we have $\vec p \succeq \vec p'$ if and only if there exists a probability distribution $q_\alpha$ of permutations $\pi^{(\alpha)}$ such that
\begin{align}\label{eq:defmajclassical}
	\vec p' = \sum_\alpha q_\alpha \pi^{(\alpha)} \vec p.
\end{align}
The Shannon entropy of a probability vector $\vec p=(p_1,\ldots,p_d)^\top$ is defined as 
\begin{align}
	H(\vec p) = -\sum_{j=1}^d p_j \log(p_j).
\end{align}
Finally, the trace-distance is replaced by the total variation distance given by
\begin{align}
	D(\vec p,\vec p') := \frac{1}{2} \sum_{j=1}^d |p_j-p_j'|.
\end{align}
A formal definition of catalytic transitions in the classical case can now be given in full analogy to the quantum case:
\begin{definition}[Classical approximate catalytic transitions] For two probability vectors $\vec p,\vec p'\in\mathbb R^{|S|}$ on a system $S$ we write $\vec p\rightarrow_\epsilon \vec p'$ if there exists a finite-dimensional probability vector $\vec q\in \mathbb R^{|C|}$ on a system $C$ and permutation matrix $\pi$ on $\mathbb R^{|S|}\otimes \mathbb R^{|C|}$ such that 
	\begin{align}
		\Tr_S[\pi \vec p\otimes \vec q]=\vec q\ \text{and}\ D\big(\Tr_C[\pi \vec p\otimes \vec q],\vec p'\big)\leq \epsilon. 
	\end{align}
\end{definition}
For completeness, we also state again the theorem from the main text we want to prove.
\begin{theorem}\label{thm:ccec2}
	Let $\vec p,\vec p' \in \mathbb R^d$ be two probability vectors with Shannon entropies $H(\vec p)$ and $H(\vec p')$, respectively. The following are equivalent:
	\begin{enumerate}
	\item For all $\epsilon>0$ we have $\vec p \rightarrow_\epsilon \vec p'$.
	\item $H(\vec p)\leq H(\vec p')$. 
	\end{enumerate}
\end{theorem}
We now prove Theorem~\ref{thm:ccec2}. The implication i) $\Rightarrow$ ii) follows as before from additivity, sub-additivity, continuity as well as invariance under permutations of Shannon entropy. We therefore only need to prove the converse direction.  
As all the results about typicality transfer unchanged (up to notation) we will freely use them in the proof. 
In particular Lemma~\ref{lemma:typmaj} implies that for sufficiently large $n$ we have
\begin{align}\label{eq:classic_maj}
	\vec p^{\otimes n} \succeq \vec p'_{\epsilon,n}
\end{align}
with $D\big({\vec p'}^{\otimes n} , \vec p'_{\epsilon,n}\big)\leq \epsilon$ and the same error scaling as before.
Let $(q_\alpha,\pi^{(\alpha)})$ be the corresponding distribution over permutations required for the majorization in \eqref{eq:classic_maj}. We again introduce $n-1$ systems $S_2\cdots S_n$, an auxiliary system $A$ with $|A|=n$ and  a source of randomness $R$ whose dimension now coincides with the number of permutations in the collection $(q_\alpha,\pi^{(\alpha)})$ (at most $d^n$ by Caratheodory's theorem). Fig.~\ref{fig:structure_catalyst_c} illustrates the structure of the catalyst.  
We denote the canonical basis vectors on $A$ by $\vec a_k$ and those on $R$ by $\vec r_\alpha$ for clarity of notation. Finally, define
\begin{align}
	\vec x^{(\alpha)}_{1,\ldots,i} := \Tr_{S_{i+1}\cdots S_n}\left[\pi^{(\alpha)} \vec p^{\otimes n}\right]
\end{align}
with $\vec x^{(\alpha)}_0=1$ the trivial state. 
We then define the catalyst state
\begin{align}
\vec q := \frac{1}{n} \sum_{k=1}^n \sum_{\alpha} q_\alpha \vec p^{\otimes k-1} \otimes \vec x^{(\alpha)}_{1,\ldots,n-k} \otimes \vec a_k\otimes \vec r_\alpha 
\end{align}
and the permutation matrix on $S_1\cdots S_n AR$ with $S=S_1$ given by
\begin{align}
	\pi := \sum_{\alpha} \pi^{(\alpha)} \otimes (\vec a_n^{\phantom\top} \vec a_n^\top) \otimes (\vec r_\alpha^{\phantom\top} \vec r_\alpha^\top) + \mathbf 1_{S_1S_2\cdots S_n} \otimes P_A \otimes \mathbf 1_R,
\end{align}
where $P_A$ denotes the orthonormal projector onto $\mathrm{span}\{\vec a_1,\ldots,\vec a_{n-1}\}$. Note that $\vec r_\alpha^{\phantom \top}\vec r_\alpha^\top$ is simply the orthonormal projector onto the span of $\vec r_\alpha$ and similarly for $\vec a_k^{\phantom \top} \vec a_k^\top$.
It is worth to take a pause to understand the structure of the catalyst state $\vec q$: If $A$ is in state $\vec a_k$ it consists of $k-1$ copies of $\vec p$ together with the first $n-k$ marginals of the state $\vec p'_{\epsilon,n}$. However, these latter marginals are correlated with the source of randomness $R$ whose marginal is such that it is in state $\alpha$ with probability $q_\alpha$. Thus, the source of randomness is correlated with all the other systems, but in a well controlled way.   
The permutation $\pi$ has the effect of applying $\pi^{(\alpha)}$ to the systems $S_1\cdots S_n$ if system $A$ is in state $n$ and system $R$ is in state $\alpha$. If these conditions are not met, it does nothing.

The final procedure now again consists of three steps (see again Fig.~3 in the main text):
\begin{enumerate}
 \item Apply $\pi$ to the initial state $\vec p \otimes \vec q$, obtaining the state
	 \begin{widetext}
	 \begin{align}
		 \pi \vec p\otimes \vec q = \frac{1}{n}\sum_\alpha q_\alpha \pi^{(\alpha)} \vec p^{\otimes n} \otimes \vec a_n \otimes \vec r_\alpha +   \frac{1}{n} \sum_{k=1}^{n-1} \sum_{\alpha} q_\alpha \vec p^{\otimes k} \otimes \vec x^{(\alpha)}_{1,\ldots,n-k} \otimes \vec a_k\otimes \vec r_\alpha
	\end{align}
	 \end{widetext}
\item Apply the cyclic shift $\mathbb S$ which acts as $S_k\mapsto S_{k+1}$ among the subsystems $S_i$. 
\item Apply the cyclic shift $\vec a_k \mapsto \vec a_{k+1}$ on the system $A$. 
\end{enumerate}
Observe that all steps correspond to permutation matrices, and therefore their composition is also a permutation matrix. 
Noting that $\Tr_{S_1}[\mathbb S\, \cdot\, ]= \Tr_{S_n}[\,\cdot\,]$, the final state on the catalyst is then given by
\begin{widetext}
\begin{align}
	&\Tr_{S_n}[\frac{1}{n}\sum_\alpha q_\alpha \pi^{(\alpha)} \vec p^{\otimes n}] \otimes \vec a_1 \otimes \vec r_\alpha +   \frac{1}{n} \sum_{k=1}^{n-1} \sum_{\alpha} q_\alpha \Tr_{S_n}[\vec p^{\otimes k} \otimes \vec x^{(\alpha)}_{1,\ldots,n-k}] \otimes \vec a_{k+1}\otimes \vec r_\alpha\\
	&=\frac{1}{n}\sum_\alpha q_\alpha \vec x^{(\alpha)}_{1,\ldots,n-1} \otimes \vec a_1 \otimes \vec r_\alpha +   \frac{1}{n} \sum_{k=2}^{n} \sum_{\alpha} q_\alpha \Tr_{S_n}[\vec p^{\otimes k-1} \otimes \vec x^{(\alpha)}_{1,\ldots,n-k+1}] \otimes \vec a_{k}\otimes \vec r_\alpha\\
	&=\frac{1}{n}\sum_\alpha q_\alpha \vec x^{(\alpha)}_{1,\ldots,n-1} \otimes \vec a_1 \otimes \vec r_\alpha +   \frac{1}{n} \sum_{k=2}^{n} \sum_{\alpha} q_\alpha \vec p^{\otimes k-1} \otimes \vec x^{(\alpha)}_{1,\ldots,n-k} \otimes \vec a_{k}\otimes \vec r_\alpha \\
	&=\frac{1}{n} \sum_{k=2}^{n} \sum_{\alpha} q_\alpha \vec p^{\otimes k-1} \otimes \vec x^{(\alpha)}_{1,\ldots,n-k} \otimes \vec a_{k}\otimes \vec r_\alpha \\
	&= \vec q. 
\end{align}
\end{widetext}
Similarly, it is easy to check that the final state on the system $S$ is given by
\begin{align}
	\Tr_{S_2\cdots S_n AR}[\pi \vec p\otimes \vec q] = \frac{1}{n}\sum_{k=1}^n \Tr_{\overline{k}}[\vec p'_{\epsilon,n}],
\end{align}
where $\Tr_{\overline{k}}$ again denotes taking the marginal of subsystem $S_k$.
Again we find
\begin{align}
	D\big(\vec p',\frac{1}{n}\sum_{k=1}^n \Tr_{\overline{k}}[\vec p'_{\epsilon,n}]\big)\leq \epsilon.
\end{align}
This finishes the proof.

\section{Size of the catalyst}
\label{sec:size}

It is interesting to ask how the construction of the catalyst scales in certain limiting cases. 
The dimension of the catalyst is essentially controlled by the number of copies $n$ required to achieve a certain precision $\epsilon$ in the construction of the catalyst,
which in turn only depends on the error for majorization from Lemma~4. This error leads to a scaling of the required number of copies $n$ as
\begin{align}
	n = O\left(\frac{\log(1/\epsilon)}{\Delta H^2}\right)
\end{align}
and hence the dimension of the catalyst diverges as $d_c = O(\exp(\log(1/\epsilon)/\Delta H^2))$.
In particular, the catalyst dimension may in general diverge in a state transformation between two states with almost the same entropy.  
In Ref.~\cite{Boes2020} a lower bound for exact catalytic transitions  was proven for state-transition with a fixed change 
of  the \emph{variance of surprisal} $V(\rho) = \tr[\rho(-\log(\rho)-H(\rho))^2]$ but a small change of entropy. 
It was found that the dimension of the catalyst has to grow at least as as $O(\exp(\Delta H^{-1/8}))$, which is compatible with the above estimate.
It is worth emphasizing, however, that these diverging catalysts sizes are only needed when $\rho$ does not majorize $\rho'$ already --- if $\rho\succeq \rho'$ the catalyst can be very small, namely of dimension $d$ in the classical case and $\sqrt{d}$ in the quantum case \cite{Boes2018a}. 
\end{document}